\documentclass[12pt,preprint]{aastex}

\slugcomment{}

\shorttitle{Binary Protostars: Analysis}
\shortauthors{Connelley et al.}

\begin{document}

\title{The Evolution of the Multiplicity of Embedded Protostars II: Binary Separation Distribution \& Analysis\footnote{The Infrared Telescope Facility is operated by the University of Hawaii under Cooperative Agreement no. NCC 5-538 with the National Aeronautics and Space Administration, Science Mission Directorate, Planetary Astronomy Program. The United Kingdom Infrared Telescope is operated by the Joint Astronomy Centre on behalf of the Science and Technology Facilities Council of the U.K. Based in part on data collected at Subaru Telescope, which is operated by the National Astronomical Observatory of Japan.}}

\author{Michael S. Connelley\altaffilmark{1}, Bo Reipurth\altaffilmark{2}, and Alan T. Tokunaga\altaffilmark{3}}

\altaffiltext{1}{NASA Ames Research Center, MS 245-6, Moffett Field, CA 94035}
\altaffiltext{2}{University of Hawai'i Institute for Astronomy, 640 N. A'ohoku Pl., Hilo HI 96720}
\altaffiltext{3}{University of Hawai'i Institute for Astronomy, 2680 Woodlawn Dr., Honolulu, HI 96822}

\begin{abstract}

      We present the Class I protostellar binary separation distribution based on the data tabulated in a companion paper.  We verify the excess of Class I binary stars over solar-type main sequence stars in the separation range from 500~AU to 4500~AU.  Although our sources are in nearby star forming regions distributed across the entire sky (including Orion), none of our objects are in a high stellar density environment.  A log-normal function, used by previous authors to fit the main sequence and T Tauri binary separation distributions, poorly fits our data, and we determine that a log-uniform function is a better fit.  Our observations show that the binary separation distribution changes significantly during the Class I phase, and that the binary frequency at separations greater than 1000~AU declines steadily with respect to spectral index.  Despite these changes, the binary frequency remains constant until the end of the Class I phase, when it drops sharply.  We propose a scenario to account for the changes in the Class I binary separation distribution.  This scenario postulates that a large number of companions with a separation greater than $\sim1000$~AU were ejected during the Class 0 phase, but remain gravitationally bound due to the significant mass of the Class I envelope.  As the envelope dissipates, these companions become unbound and the binary frequency at wide separations declines.  Circumstellar and circumbinary disks are expected to play an important role in the orbital evolution at closer separations.  This scenario predicts that a large number of Class 0 objects should be non-hierarchical multiple systems, and that many Class I YSOs with a widely separated companion should also have a very close companion.  We also find that Class I protostars are not dynamically pristine, but have experienced dynamical evolution before they are visible as Class I objects.  Our analysis shows that the Class I binary frequency and the binary separation distribution strongly depend on the star forming environment.  

\end{abstract}

\keywords{binaries: general, stars: formation, stars: statistics, infrared: stars}

\section{Introduction}

   It is well known that most stars are in binary star systems, so understanding the problem of star formation requires an understanding of binary star formation.  \citet{Duq1991} found that the binary frequency of solar-type main sequence stars is $\sim$60\%, with a peak in the binary separation distribution at 30~AU.  The multiplicity of pre-main sequence and T Tauri stars has been investigated at a range of wavelengths using a variety of imaging techniques (e.g. Ghez et al. 1993, Leinert et al. 1993, Reipurth \& Zinnecker 1993, Simon et al. 1995, K\"{o}hler \& Leinert 1998, Padgett et al. 1997, Beck et al. 2003, Ratzka et al. 2005).  T Tauri stars are found to have an excess of binary companions relative to main sequence stars (62\% \citep{Pat2002} vs. 46\% \citep{Duq1991} from 0.02~AU to 560~AU).  \citet{Pat2002} also found that the mean separation of T Tauri companion stars is 130~AU.  
   
   Class I young stellar objects (YSOs) are characterized by a positive spectral index \citep{Lad1991}, when measured from $\sim10~\mu$m to $\sim100~\mu$m, and the presence of a molecular envelope that is typically less massive than the central star \citep{Bon1996}.  They are believed to be $\sim10^5$ years old, or roughly 10 times younger than average T Tauri stars.  \citet{Hai2004} observed that the Class I binary frequency is 18\%$\pm$4\% in the separation range from 300~AU to 2000~AU.  \citet{Duc2004} found a binary frequency of 27\%$\pm$6\% in the separation range from 110~AU to 1400~AU, and also found a possible correlation between the frequency of widely separated binary companions and the presence of an extended millimeter envelope.  They show that Class I protostars have a binary excess over solar-type main sequence stars but have a binary frequency consistent with T Tauri stars.  The decreasing binary frequency from the T Tauri phase to the main sequence, as well as the main sequence binary properties led Larson (2001) to conclude that the vast majority of stars form in binary or multiple systems and not in isolation.  Observations by \citet{Duc2004} also suggest that the binary frequency may change during the Class I phase. 
   
   Pre-main sequence binary properties have also been investigated in different star forming environments.  The binary frequency has been found to be lower in higher density star clusters \citep{Pat2001}.  T Tauri stars in the Chamaeleon, Lupus, and Corona Australis star forming regions have fewer companions wider than 125~AU but more companions closer than 125~AU than T Tauri stars in the Taurus and Ophiuchus star forming regions \citep{Ghe1997}.  The $\eta$ Chamaeleontis group also has a dearth of widely separated companion stars relative to the TW Hydrae group, which has a similar age \citep{Bra2006}.  \citet{Rei1993} found a greater number of T Tauri binary stars in clouds with fewer than 10 stars than in clouds with more than 10 stars.  However, \citet{Duc2007} conclude that the data on the binary properties of Class I YSOs show no clear dependence on the star forming region or environment.  

   The following analysis is based on the data presented in Connelley, Reipurth, \& Tokunaga (2008, in press) herein Paper I.  That paper presents observatons of a new IRAS selected sample of 189 Class I YSOs distributed across the whole sky north of $\delta\approx-40^{\circ}$, including the Taurus, Ophiuchus, and Orion star forming regions (but not in the Orion Nebula Cluster). Those observations found 78 binary companions in the separation range from $\sim$100~AU to 5000~AU and with a contrast less than $\Delta$L$'$=4 magnitudes.   In this paper, we verify the Class I binary frequency results of \citet{Duc2004} and \citet{Hai2004} with a larger and more diverse sample of Class I YSOs observed with higher angular resolution, and present a binary separation distribution for Class I YSOs for comparison to the solar-type main sequence and T Tauri binary separation distributions.  We further explore how the binary separation distribution evolves during the Class I phase, and propose a scenario to explain part of this.  Finally, we consider the dependence of the Class I binary statistics on the star forming region.    
   
\section{Binary Separation Distribution}
\subsection{Conversion to Semi-major Axis}

    The observational results presented in Paper I include the distances to the clouds that host the IRAS sources and the projected separations of the companions.  However, in order to construct the binary separation distribution, we need the actual semi-major axis of the companion star's orbit. \citet{Kui1935} empirically showed that the ratio of the average projected separation ($\rho$) to the average semi-major axis (\emph{a}) for a large sample of binary orbits is $\rho/a=0.776$.  This scaling relation was used to convert our projected separations to an average value for the semi-major axes of the orbits.  We note that most other studies that we compare our results to have not used this conversion.  We compare our results with those from \citet{Duq1991}, which are given in terms of orbital period.  Since they observed a sample of main sequence stars with spectral types ranging from F7 to G9, we used Kepler's $3^{rd}$ Law with a system mass of 1.2~M$_{\odot}$ to convert from orbital period to orbital semi-major axis.   

 \subsection{Method for Incompleteness Correction}
      The separation and contrast space where binary companions could be found was divided into a grid of separation and contrast bins.  We defined twelve bins of separation log(\emph{d}/1~AU) = $0.1\bar{6}$ wide (one fourth of the width used by Duquennoy \& Mayor 1991).  The contrast bins are 1 magnitude deep, from $\Delta$L$'=$~0 to 1, 1 to 2, 2 to 3, and 3 to 4.  The binary companion counting and incompleteness correction in each bin of separation and contrast were done independently of all other separation/contrast bins.  For the incompleteness correction, we used the inner detection limit for the higher contrast end of each contrast bin, e.g. for the contrast range from $\Delta$L$'=$~1 to 2, we used the inner detection radius for companions $\Delta$L$'=$~2 magnitudes fainter than the primary star.  

      The incompleteness correction was done as follows.  We assume that the undetected binaries have the same properties as the ones we did detect.  For each bin of separation and contrast (e.g. from 207~AU to 305~AU and from $\Delta$L$'=$~1 to 2), we counted the number of companion stars found in that bin as well as the number of targets for which we \emph{could} have found a companion in that bin (\# sensitive), using the inner and outer separation limits tabulated in Paper I.  If we were sensitive to companion stars for a fraction of the separation range, then this was counted as that fraction of a star.  These numbers were used with the total number of stars in the survey (\# total) to determine the number of stars for which we could not detect a companion (\# insensitive = \# total $-$ \# sensitive) within this separation and contrast bin.  This was then used to determine the correction factor (\# insensitive / \# sensitive) by which the number of observed companions was multiplied to get the estimated number of missed companions.  The counting uncertainty of the number of observed companions was also multiplied by this correction factor to determine the uncertainty in the number of missed companions.  The number of found and the number of missed companions (along with their errors) were added together to get an incompleteness corrected total number of companions for that bin of separation and contrast.  Adding the incompleteness corrected numbers of companions together for all four contrast bins yielded the incompleteness corrected number of companions for that bin of separation.  

\subsection{A Word About Statistics}
     We adopted Poisson statistics to determine the 1~$\sigma$ uncertainties of our measurements.  In order to determine how similar or different our separation distributions are from previous results, we compared the results from each of our separation bins with the result from the matching bin from a comparable study (for example, Duquennoy \& Mayor 1991).  We used the equation given by \citet{Bra2006} in their appendix B2 to calculate the probability that the results in a given separation bin from two studies are consistent with each other.  This equation uses binomial statistics and allows for different sample sizes.  We multiplied together this result from each of the separation bins to determine the probability that the two distributions are consistent with each other.  We did not use the Kolmogorov-Smirnov test to determine if two binary separation distributions are consistent with each other, since the data are not sensitive to binaries of all separations.  Also, the K-S test cannot distinguish distributions that are scaled versions of the same function.  

\section{Comparisons With Previous Work}
\subsection{Main Sequence and T Tauri stars}
    Two studies that summarize the binary separation distributions for solar-type main sequence stars and T Tauri stars are \citet{Duq1991} and \citet{Pat2002}.  \citet{Pat2002} found that the companion star frequency is 62\% from 0.02~AU to 560~AU for T Tauri stars in the Taurus, Ophiuchus, Chamaeleon, Corona Australis, and Lupus clouds, and is 14.5\% from 100~AU to 560~AU (a separation range that overlaps with our data).  The studies they cite have sample sizes ranging from 69 to 254.  \citet{Duq1991} observed a sample of 164 dwarf stars with spectral types ranging from F7 to G9.  Their incompleteness corrected binary frequency is $\sim$60\% integrated over all separations.  Interpolating their data over the separation range from $\sim100$~AU to 4500~AU (a separation range that also overlaps with our data), the solar type main sequence binary frequency is 16\%.   In comparison, our Class I binary frequency from $\sim100$~AU to 4500~AU is 43\%. In the separation range from 100~AU to 560~AU, our Class I binary frequency is 18\% versus 14.5\% for T Tauri stars in the same separation range.  Within the separation ranges stated above, we see an excess of Class I binary companions versus solar-type main sequence stars, whereas the binary frequencies of Class I and T Tauri stars are very similar.  \citet{Duq1991} fit their binary separation distribution with a log-normal function (i.e. a Gaussian on a log scale).  Scaling the same log-normal distribution function to fit our data would result in a binary frequency of 180\%, integrated over all separations.  Clearly, this result is unreasonable.  Thus, \emph{the binary separation distribution function must change from the Class I phase to the main sequence.} 

  Both \citet{Duq1991} and \citet{Pat2002} fit their binary separation distributions with a log-normal function.  \citet{Kui1955} also noted that the main sequence binary separation distribution is roughly Gaussian in log(\emph{a}), and used angular momentum arguments to derive a distribution with a log-normal profile that peaks at 25~AU.  On the other hand, \citet{Opi1924} suggested using a log-uniform function (i.e. a function that has a constant value versus log-separation) to fit the main sequence binary separation distribution as it was known at the time.  To test whether a log-normal or a log-uniform function is a better fit to the Class I binary separation distribution, we measured $\chi^{2}$ between our data points and the log-normal function used by \citet{Duq1991} and a log-uniform function, each scaled to minimize the value of $\chi^{2}$.  The minimum $\chi^{2}$ for the scaled log-normal function is 9.04 and for the scaled log-uniform function is 1.56, thus the log-uniform function is a much better fit.  Since we cannot determine what function describes the Class I binary separation distribution over all separations, we cannot fit a function to our data to extrapolate the Class I binary separation distribution to separations closer than 50~AU.  As such, we cannot estimate the Class I binary frequency integrated over all separations.

  Figure 1 shows our result overlaid with the result from \citet{Duq1991}, with their result scaled to match our separation bin size.  Using the confidence analysis from \citet{Bra2006} described above to compare our result with the result from \citet{Duq1991}, the probability that Class I YSOs have the same binary separation distribution as solar-type main sequence stars is $2.06\times10^{-5}$.  At separations wider than 100~AU, we observed a binary frequency greater than the solar-type main sequence binary frequency.  However, the difference is greatest beyond 500~AU (log(\emph{d}/1~AU)=2.7), where the Class I binary frequency begins to increase.  The Class I and T Tauri binary separation distributions (Figure 2) are consistent to separations as wide as 2000~AU.  

   The slight rise in the binary frequency at wide separations seen in Figure 1 is noteworthy.  It is not due to an error in the application of the incompleteness correction, since the incompleteness correction is smallest at wide separations.  Contamination is immediately suspected as the cause of this rise.  As detailed in Paper I, we adopted an outer separation limit that limits the chance of field star contamination to an average of 3.0\% (i.e. we expect that 6 identified companions are spurious).  If these 6 contamination stars are distributed among the outermost three separation bins, then the third to the last bin would have 0.77 stars, the second to last bin would have 1.56 stars, and the outermost bin would have 3.56 stars.  When these numbers of stars are removed from these bins, the slight rise at wide separations is diminished, as shown in the figure.  This contamination correction has been adopted in subsequent figures.

\subsection{Previous Class I studies}
   Two studies of the Class I binary frequency were presented by \citet{Duc2004} and \citet{Hai2004}.  \citet{Duc2004} observed 63 Class I stars at K-band in the Taurus and Ophiuchus clouds and found a binary frequency of $27\%\pm6\%$ from 110~AU to 1400~AU.  \citet{Hai2004} observed 76 Class I stars at K-band in the Taurus, Ophiuchus, Serpens, Perseus, Chamaeleon I and II clouds and found a binary frequency of $18\%\pm4\%$ from 300~AU to 2000~AU.  Figure 3 shows our results overlaid with the results from \citet{Duc2004} and \citet{Hai2004}, with their results scaled to match our bin size.  Despite the differences between the samples among these three studies, the results are consistent with each other.  However, as shall be shown below, this does not mean that the binary frequency is independent of the star forming regions that are observed.

\section{Higher Order Multiple Systems}

    In addition to binaries, many systems are triple or quadruple stars.  \citet{Tok2004} found that 15\% to 25\% of all F through K-type stars are part of higher order multiple systems, and that nearly all stars with a close binary companion also have a wide companion.  Our survey of Class I protostars also found many triple and quadruple star systems.  We consider a star to be a triple or quadruple if all companions have a projected separation less than 5000~AU from the primary star.  In the case of the quadruple systems, the stars of IRAS 17369$-$1945 are in a close apparently non-hierarchical group whereas IRAS 18270$-$0153 is a hierarchical double binary star system.   In our sample of Class I YSOs, the numbers of single, binary, triple, and quadruple stars are S:B:T:Q=122:51:12:4, which is S:B:T:Q = 2.39:1.00:0.24:0.08 when normalized to the number of binaries.  In a survey of 52 visual pre-main sequence binary stars, \citet{Cor2006} found that the number of binaries to triples to quadruples is 30:5:6 (1.00:0.17:0.20 normalized to the number of binaries) in the separation range from 10~AU to 2300~AU. Their finding of a large number of quadruples may be due to their higher spatial resolution ($\sim$80~mas).  However, we find a similar fraction of higher order multiples versus binary stars ($31.4\%^{+7.8\%}_{-7.0\%}$ (16/51) for our study, $36.7\%^{+10.7\%}_{-9.7\%}$ (11/30) for \citet{Cor2006}).
   
   
   Considering binary companions with periods greater than 10$^{5}$ days ($\sim$50~AU), the numbers of single, binary, triple, and quadruple stars observed by \citet{Duq1991} is S:B:T:Q=131:32:1:0, which is S:B:T:Q = 4.09:1.00:0.03:0.00 normalized by the number of binaries.  We observed that $\sim24\%$ (16/67) of our non-single protostellar systems are higher order multiples, whereas only $\sim3\%$ of non-single stars are higher order multiples among solar-type main sequence stars at separations wider than 50~AU.  For separations greater than 50~AU, Class I protostars are 1.4 times more likely to be binaries and 13.9 times more likely to be higher order multiples than solar-type main sequence stars.  

\section{Spectral Index Dependence}
\subsection{Spectral Index as a Proxy for Age}

    In order to examine how the binary frequency distribution changes within the Class I phase, we needed a parameter that traces the relative age of all of our YSOs.  While other age tracers, such as the presence of an extended millimeter envelope \citep{Duc2004}, have been used, we used spectral index ($\alpha$).  Spectral index\footnote{Different definitions of spectral index have been used in the literature.  We adopt the convention set by \citet{Lad1991}, where the spectral index $\alpha=-$dlog$\nu$F$_{\nu}$/dlog$\nu$.  In this convention, Class 0 and I YSOs have a positive spectral index, whereas Class II and III objects have a negative spectral index from $\sim10~\mu$m to $\sim100~\mu$m.  We used the IRAS fluxes from 12~$\mu$m to 100~$\mu$m, or from 25~$\mu$m to 100~$\mu$m if the source was not detected at 12~$\mu$m.} is the only age tracer available for every target since our sample was selected from the IRAS catalog.  A source with a higher spectral index tends to have a more massive cold envelope and comparatively less warm circumstellar dust than a source with a lower spectral index.  Observations of Class I reflection nebulae showed that the amount of circumstellar material, traced by the size and brightness of the nebula and how clearly the central source is seen at K-band, correlates with the spectral index \citep{Con2007}.  However, we do not know how to directly convert spectral index into age.  Since earlier evolutionary phases are expected to have shorter durations, YSOs should pass through the higher spectral index phases quickly and spend more time with a lower spectral index, in which case a conversion between spectral index and time would be non-linear.   Spectral index is not an ideal relative age tracer.  For example, an edge-on disk can give a more evolved YSO the spectral index of a younger star.  Nevertheless, spectral index is the best relative age tracer available for all of our sources.

\subsection{The Evolution of the Binary Separation Distribution}

     To explore how the binary separation distribution changes with time, we divided our sample into four groups of roughly equal size (48 to 51 IRAS sources each) based on their spectral indices.  Our goals were to determine if the binary frequency and the binary separation distribution change within the Class I phase.  Figures 4 and 5 show the binary separation distribution in 4 domains of spectral index. These figures each have three bins of separation log(\emph{d}/1~AU)=$0.6\bar{6}$ wide.  The accounting and incompleteness correction were done using log(\emph{d}/1~AU)=$0.1\bar{6}$ wide separation bins, the results of which were added together to derive the result for each log(\emph{d}/1~AU)=$0.6\bar{6}$ wide bin. The least evolved domain includes objects with spectral indices from 1.09 to 2.16.  Within this domain of spectral index, nearly all of the companions have separations greater than 1000~AU.   The second domain includes objects with spectral indices from 0.66 to 1.09.  Here, the binary frequency at separations wider than 1000~AU is much lower than in the previous spectral index domain yet the binary frequency at separations closer than 150~AU is quite high.  The third domain includes objects with spectral indices from 0.36 to 0.66.  Overall, this binary separation distribution is very similar to the previous spectral index domain.  The fourth and presumably most evolved domain includes objects with a spectral index less than 0.36.  The binary frequency at separations greater than 1000~AU is a third of the value in the previous domain and no companions were found at separations less than 150~AU.  In this last spectral index domain, the binary frequency is $16.5\%\pm5.5\%$, whereas the solar-type main sequence binary frequency is $23.2\%\pm3.8\%$ over the same separation range.  

   The overall decrease in the separation of the binaries is shown in Figure 6.  The Class I binary separation is observed to decline with respect to decreasing spectral index throughout the Class I phase.  The median separation does not appear to change in the last two spectral index domains, although the overall binary frequency changes greatly between these index domains, as shown in the next figure.  The error bars are large since each spectral index domain still has binary stars with a wide range of separations.  These error bars are the median standard deviation of the separations in each spectral index domain.  Instead of using the square root of the \emph{average} squared deviation from the mean value, we used the square root of the \emph{median} squared deviation from the median value.  This form of the standard deviation is less affected by outlying data points than the form of the standard deviation that is typically used. 

   We used the binomial confidence test in appendix B of \citet{Bra2006} to determine if the evolution of the binary separation distribution we observed with respect to spectral index is statistically significant.  The probability that the binary separation distributions of the first and second spectral index domains are consistent with each other is $3.3\times10^{-7}$.  This probability for the second and third spectral index domains is $3.6\times10^{-2}$, and for the third and fourth spectral index domains is $7.5\times10^{-5}$.  The probability that the binary separation distributions of the first and fourth spectral index domains are consistent with each other is $9.16\times10^{-9}$.  We are therefore highly confident that \emph{there is a statistically significant change in the binary separation distribution within the Class I phase}.

   Is the dearth of close and very young binary companions the result of close companions lying hidden behind a disk or being more deeply embedded in the circumstellar envelope?  It is possible that a close companion may be obscured by the circumstellar disk of the primary star, especially during the earlier parts of the Class I phase when the disk and the envelope are the most massive.  If a significant number of close companions are being hidden by an opaque disk, we would expect that YSOs with observable companions closer than 200~AU (roughly the size of the disk) would tend to have a lower spectral index (i.e. be more evolved) than sources with a more widely separated companion.  We used the two sample Kolmogorov-Smirnov test to compare the spectral index distributions for the 19 sources with companions separated by less than 200~AU and the 60 sources with companions separated by more than 200~AU.  The probability that these distributions are different is less than 90\%, thus these distributions are statistically consistent with each other.  We conclude that a significant number of close companions are not being obscured by an opaque circumstellar disk.  

   Why would companions closer than 200~AU not be obscured by a circumstellar disk?  If a companion star has a separation less than 200~AU, its orbital period is much less than the Class I life time.  As such, the two circumstellar disks should be truncated to a radius significantly less than the periastron separation.  However, the companion is most likely to be observed near apastron.  Thus, it is unlikely that a companion star would be hidden from our line of sight by the primary star's circumstellar disk due to the truncation of the primary star's circumstellar disk by the companion.
 
     Figure 7 shows the change in the binary frequency as a function of spectral index for all companions with separations from $\sim50$~AU to $\sim4500$~AU.  Since the spectral index ($\alpha$) decreases with time, YSOs evolve towards the right of the figure.  The binary frequency remains relatively constant at $\sim55\%$ from $\alpha>2$ to $\alpha\sim0.5$, despite the fact that the binary separation distribution changes significantly during this time.  For $\alpha<0.5$, the binary frequency experiences a $4.0\sigma$ drop to less than a third of its previous value.  The decline in the binary frequency late in the Class I phase is likely due to several combined effects.  The binary frequency at wide separations declines steadily relative to spectral index (see Figures 4 and 5).  The binary frequency at close separations also declines late in the Class I phase, possibly due to orbital migration to separations too close for our data to resolve (for further discussion, see Section 5.5).  

    We also considered how the binary frequency in each separation bin changes with spectral index.  The evolution of the binary frequency at separations from 45~AU to 207~AU is shown in Figure 8.  The binary frequency is initially very small, increases to a value of $\sim15\%$ for two spectral index bins, then drops to a very low value in the most evolved spectral index bin.  At separations from 207~AU to 963~AU (Figure 9), the binary frequency is $\sim5\%$ in the two youngest spectral index domains.  It increases to 23\% in the third index domain then drops to 11\% for the oldest spectral index domain.  At wide separations (963~AU to 4469~AU), as shown in Figure 10, the binary separation declines at a steady rate with respect to the spectral index.  A linear fit to the data is overlaid on the data, and has the form \emph{bf} = $-$0.02 + $0.36 \cdot \alpha$, where \emph{bf} is the binary frequency and $\alpha$ is the spectral index.  The fact that the binary frequency at this separation declines at a constant rate with respect to the spectral index, and the fact that this decline occurs throughout the whole Class I life time, are both very important clues to the mechanism behind this evolution.

\subsection{Theoretical Considerations: Making Binaries}

    The three processes that have been examined for the formation of binary stars are fragmentation, capture, and fission.  Fission is no longer considered a viable mechanism for binary star formation, since the development of a bar instability creates spiral arms that dissipate angular momentum \citep{Dur1986}, preventing the star from breaking into separate stars.   Capturing two unbound stars into a bound binary requires a dissipative medium, such as a circumstellar disk, stellar tides, or a third star.  A very high stellar density is required for any of these three media to be efficient, and thus these mechanisms are not likely to facilitate the capture of a binary companion \citep{Bon2001}.  Furthermore, since none of our Class I objects are in a dense stellar environment, capture is not an effective means of forming the young binary stars that we observed.

    Fragmentation of a dense core, either due to rotation or turbulence, is the favored means of binary star formation.  The expected mean separation for a binary star formed by fragmentation is $125(M_{core}/M_{\odot})^{1/3}$~AU \citep{Goo2007}.  Binary stars made via rotationally driven fragmentation are expected to have a similar mean separation of $130(\alpha_{o}/0.5)(\beta_{o}/0.02)(M/M_{\odot})(10K/T)$~AU \citep{Ste2003}, where $\alpha_{o}$ is the ratio of the thermal to gravitational energy and $\beta_{o}$ is the ratio of rotational to gravitational energy.  However, we have observed that the youngest Class I binary stars have separations much wider than a few$\times$100~AU (as predicted by the equation above, substituting $\alpha_{o}\approx0.5$, $\beta_{o}\approx0.02$, M$\approx$2~M$_{\odot}$, and T$\approx$10~K).  Therefore, either most binary stars do not form via fragmentation, or the theoretically derived mean separations are incorrect, or the properties of Class I binary stars do not reflect the properties of binary stars as they initially formed.

   Fragmentation within a massive circumstellar disk is also a potential means of making a companion star.  A numerical simulation by \citet{Bon1994_837} has shown that the infall of gas allows spiral arms in the disk to gain a Jeans' mass, and thus collapse into a companion star.  Numerical simulations by \cite{Bur1997} follow a 1~M$_{\odot}$ cloud as it evolves into a disk and fragments.  The arrangement of these fragments spans less than 200~AU and is dynamically unstable, thus they anticipate there would be multiple ejections as the system rearranges itself into a hierarchical multiple star.  \citet{Bon1994_999} also find that the interaction of a central binary star with its circumbinary disk can trigger the fragmentation and collapse of a new companion in the disk.  \citet{Goo2007} state that star-disk interactions play an important role in the formation of binary stars by stimulating the fragmentation of the disk if the disk is more than 0.5~M$_{\odot}$ in the Class 0 phase.   A recent study by \cite{Whi2006} concludes that companions can collapse from a disk at distances greater than about $150(M_{*}/M_{\odot})^{1/3}$~AU with a minimum mass of $0.003(M_{*}/M_{\odot})^{-1/4}(L_{*}/L_{\odot})^{3/8}$~M$_{\odot}$.  Due to the angular momentum of the infalling gas, it will tend to be deposited onto the disk at a radius greater than 300~AU for M$_{*}$=1~M$_{\odot}$.  They note that disks this large are rarely observed, and propose that they are converted into stars on a time scale of only 10$^{4}$ years.  They further conclude that "it is possible that close, low-mass binaries are formed in the outer parts of massive circumstellar disks".

\subsection{Theoretical Considerations: Orbital Evolution and Ejections}

    The review by \citet{Goo2007} divides dynamical evolution into two categories: 1) the dynamical decay of non-hierarchical multiple systems via ejection \citep{Rei2000} and 2) the tidal stripping of loosely bound companions in a dense cluster environment \citep{Kro1995b}.  Both of these mechanisms only decrease the companion star frequency, each with its own characteristic half-life for multiple systems. Tidal stripping is not an important process for our Class I YSOs because our sources are not in dense cluster environments and the time scale for the decline in the binary frequency (a few $\times$ 10~Myr, Kroupa 1995a) is much greater than the Class I lifetime ($\sim10^5$ years).  Small-N non-hierarchical systems\footnote{Small-N here refers to a system with more than two but less than a dozen members.} are expected to have a half-life against dynamical decay of $14R^{3/2}(M_{stars}/M_{\odot})^{-1/2}$ years, where $M_{stars}$ is the mass of the components and $R$ is the size of the system in AU \citep{Ano1986}.  This time scale is a few $\times~10^4$ years, when substituting $R\approx$150~AU and $M\approx$3~M$_{\odot}$.  The time scale for a star to gain most of its mass (i.e. to go from Class 0 to I) is also of order $10^{4}$ years.  Thus, most of the ejections are expected to happen during the Class 0 phase \citep{Rei2000}.

  If the ejection formation hypothesis of brown dwarfs \citep{Rei2001} is viable, and given the fact that brown dwarfs are not rare, ejections should be common events. There should be events where the velocity of the ejected companion has exceeded the escape velocity of the ejecting binary, but has not exceeded the escape velocity of the central binary \emph{and} its envelope.  Companions that are ejected and nearly escape solve the problem of creating the large population of very young widely separated binary companions shown in the left panel of Figure 4.  The gravity of the envelope helps to prevent the ejected star from escaping immediately.   The gravitational force on the ejected protostar thus does not decrease as $r^{-2}$, and depending on the density of the envelope as a function of radius, the gravitational force on the ejected companion can actually \emph{increase} with distance.    When ejected, such a ``nearly escaped'' star is in a highly eccentric long-period orbit but is not unbound.  Future work may determine if the escape velocity of the typical binary is less than the typical ejection velocity (3-4~kms$^{-1}$) calculated by \citet{Ste1998}, and if the escape velocity of the binary plus the envelope is greater than this value.  
 
     The expectation that non-hierarchical multiple systems dynamically decay in the Class 0 phase is supported by our finding that our youngest Class I binary stars are slightly \emph{less} likely to be in a resolvable triple system than older Class I objects.   If binaries actually formed as we see them in the Class I phase (i.e. with separations greater than 1000~AU rather than $\sim$100~AU as expected), then we would expect to see a large number of very young non-hierarchical higher order systems with separations of $\sim1000$~AU.  Such systems would be easily resolved but, with the possible exception of IRAS 17369-1945, are not found in our sample.  We conclude that the large number of young binary companions with separations greater than 1000~AU are the product of dynamical decay occuring before these stars become Class I objects, and thus Class I YSOs are not dynamically ``pristine''.

    The fact that the mean separation of solar-type main sequence and T Tauri binary stars ($\sim$30~AU (Duquennoy \& Mayor 1991) and $\sim$62~AU (Patience et al. 2002), respectively) is less than the expected formation radius ($\sim130$~AU, Goodwin et al. 2007) due to fragmentation shows that viscous interactions must play an important role in the evolution of the binary separation distribution \citep{Ghe2001}.   Interactions with gas, such as a binary star within a circumbinary disk, can reduce the semimajor axis of the binary's orbit on a time scale of $\sim10^{4}$ times the orbital period (Artymowicz et al. 1991).  \citet{Bat2002} find that tidal interactions between a central binary/multiple star and its circum-multiple disk are very efficient, i.e. a disk with a mass of $\sim10^{-2}$~M$_{\odot}$ can significantly change the orbits of the stars in the disk \citep{Pri1991}.  All of these interactions with gas can destabilize a hierarchical multiple system, leading to dynamical decay after the Class 0 phase.

\subsection{A Scenario for the Evolution of the Binary Separation Distribution}

     The binary frequency at separations wider than $\sim1000$~AU drops steadily with respect to the spectral index, declining from a binary frequency of 53\% to 5\% in an amount of time that is comparable to the Class I life time (Figure 10).  It is important to note that the decline in the binary frequency is with respect to the spectral index, which traces the distribution of material in the envelope and disk \citep{Lad1991}.  It therefore appears that the binary frequency at wide separations is correlated with the envelope mass, suggesting that the decline in the binary frequency at wide separations is linked to the loss of the envelope.  \citet{Duc2004} also found a possible correlation between the binary frequency at wide separations and the presence of an extended millimeter envelope.  We expect that a significant amount of the binding energy of a binary companion with a separation in excess of 1000~AU will be provided by the envelope, especially in the early part of the Class I phase when the envelope mass is still comparable to the mass of the primary star.  As the envelope mass decreases, the gravitational binding energy also decreases and these widely separated companions become unbound.  This scenario, as well as the fact that we are considering binary companions with separations as great as 5000~AU, requires that we are confident that these widely separated companions are likely to be gravitationally bound.  Arguments to support this are presented in Paper I.  

   This scenario requires that a companion must be lost beyond our 5000~AU search radius in much less than the Class I life time if it becomes unbound.  A star would need to move at a velocity of at least 0.24~kms$^{-1}$ to travel beyond our 5000~AU separation limit within 10$^5$ years.  For comparison, a companion in a circular orbit around a 1~M$_{\odot}$ star with a semimajor axis of 1000~AU has a velocity of 0.94~kms$^{-1}$, or a velocity of 0.54~kms$^{-1}$ for a radius of 3000~AU.  Thus, if a star in such an orbit becomes unbound, it can travel beyond our 5000~AU separation limit within the Class I lifetime.

   The binary frequency in the separation range from 45~AU to 207~AU rises from 0\% to $\sim15\%$, then drops back down to 0\% within the Class I life time (Figure 8).  The decline in the binary frequency at later times is likely to be due to viscous interactions between the binary stars and circumstellar gas, causing the separation to decrease to the point where the binary is too close for our observations to resolve.  Infalling gas and binary star/circumbinary disk interactions tend to cause the semimajor axis of the orbit to decrease \citep{Bat2002}.  Although there is expected to be less gas in the disks at these later stages of the Class I phase, it has been shown that little mass is needed in the disk to affect the orbit of a binary star \citep{Pri1991}.  

   The increase in the binary frequency within this separation range during the earliest part of the Class I phase can have several causes.  The two that are the most likely to be important are: 1) orbital migration reducing the semi-major axis of a widely separated companion, and 2) new binary companions being made \emph{in situ} by disk fragmentation.  Regarding (1): Viscous interactions between two stars with circumstellar disks or between a central binary and a circumbinary disk can harden the orbit of a binary to closer separations \citep{Pri1991}.   Reducing the semi-major axis of the orbit from more than 207~AU to less than this with a circumbinary disk would clearly require a disk that is much larger than 207~AU.  Regarding (2): Numerical simulations have shown that new binary companions can be made \emph{in situ} by disk fragmentation (Bonnell 1994, Bonnell \& Bate 1994a, Whitworth \& Stamatellos 2006, Burkert et al. 1997) at the separations where we see a rise in binary companions.  The formation of new companions can offset the losses of widely separated companions, helping to keep the binary frequency flat for most of the Class I phase (see Figure 5).  However, there may not be enough mass in the envelope and disk during the Class I phase to make new companions.  

      Our scenario for the evolution of the binary separation distribution needs two important ingredients in order for it to work: ejections and gas.  The hardening of the orbit of companions to closer separations due to star-disk interactions requires the presence of gas in the form of a circumstellar disk.  The formation of new companions in the Class I phase requires a rather massive circumstellar disk at a separation of $\sim100$~AU where it appears that new companions may be created \citep{Bur1997}.  
      
   The envelope mass should be at least comparable to the stellar mass early in the Class I phase, so that companions ejected in the Class 0 phase are kept from becoming unbound by the envelope until the envelope has dissipated.  Since the envelope of a Class 0 object is more massive than the protostar whereas the Class I protostar is more massive than its envelope (Andr\'{e} et al. 1993; Bontemps et al. 1996), it is reasonable to expect that the mass of the envelope will be comparable to the mass of the protostar during the transition between the Class 0 and I phases.  Our scenario also requires that ejections should frequently have enough energy to reach the escape velocity of the central binary but not the escape velocity of the central binary plus the envelope.   Under this condition, the ejected companion can be retained by the gravity of the envelope and central binary, then become unbound when the envelope dissipates. 

   This scenario makes numerous observationally testable predictions.  1) Class 0 objects should have a companion star fraction greater than Class I objects.  Class 0 objects should also have a higher proportion of triple and non-hierarchical multiple systems than is observed in the Class I phase.  2) Class I objects with a widely separated (and presumably ejected) companion should be more likely to have a primary star that is itself a close binary than a Class I YSO without a widely separated companion.  This has already been observed for T Tauri stars \citep{Cor2006}.  3) Tidal interactions between multiple stars and their disks should leave observable signatures in the disk \citep{Art1991}, such as disk truncation, gaps, and spiral arms \citep{Bon1994_999}.  4) Spectroscopic binaries may be rare among Class I objects, especially early in the Class I phase.  It is possible for a binary with an initial separation of $\sim100$~AU to have its orbit hardened via an ejection to $\sim10$~AU, perhaps becoming an FU Orionis object \citep{Rei2004} in the process.  Orbital migration can then reduce the period of the orbit of the binary from many years to days.  However, the reduction of the orbital period from several years to days may not be able to happen entirely within the Class I life time.  5) Orbital hardening should lead to an observable increase in the binary frequency at separations less than 50~AU late in the Class I phase.

\section{Dependence on Star Forming Region}

     There have been conflicting results regarding the dependence of binary statistics on the star forming environment.  \citet{Rei1993} found that pre-main sequence stars in clouds with 10 or fewer stars were twice as likely to have a binary companion as young stars in a cloud with more than 10 stars.  A dearth of companions at wide separations in the $\eta$ Chamaeleontis group relative to another association, the similarly aged TW Hydra group, has led \citet{Bra2006} to conclude that this discrepancy provides ``strong evidence for multiplicity properties being dependent on environment''. Most recently, \citet{Kra2007} found that the binary frequency among wide ($>$1\arcsec) binaries in the Taurus, Auriga, Chamaeleon I, and Upper Scorpius star forming associations is dependent on both the mass of the primary star and the environment (but not the stellar density).  In comparison, \cite{Duc2007} found that their data on flat-spectrum and Class I objects is consistent with there being no dependence on the star forming region.  

    In order to determine if the binary statistics of Class I YSOs have a regional or environmental dependence, we divided our sample as a whole into three groups.  Group I includes objects in Orion (i.e. all sources from 5h to 6h), Group II includes objects in nearby low stellar density star forming regions (Taurus, Auriga, Perseus, and Ophiuchus), and Group III includes all other objects not in the two previous groups (median distance is 700~pc).  We stress that our objects in the Orion region are isolated stars that are dispersed throughout the Orion molecular clouds and are \emph{not} near the Orion Nebula Cluster.  The binary separation distributions were calculated the same way as described above, and they are presented in Figures 11 and 12.  Using background star counts, we calculate that the average chance of contamination per star within a projected separation of 5000~AU and within a contrast of $\Delta$L$'$=4 in Group I is 3.4\%, for Group II is 2.7\%, and for Group III is 3.4\%, thus the expected numbers of false companions are 1.9, 1.3, and 2.3, respectively.  The figures have been corrected for this expected contamination.  The binary separation distribution for our objects in Group I has a high binary frequency at very close and very wide separations, with a relatively small number of companions in between.  We do not believe the rise in the binary frequency at wide separations is due to contamination or poor incompleteness correction since the incompleteness correction is smallest at wide separations and the chance of contamination is small and has been accounted for (for more details, see Paper I).  The binary separation distribution of our objects in Group II is comparably flat.  The binary separation distribution is also flat for targets in Group III, and is very similar to the binary separation distribution for the sample as a whole.  The binary separation distributions in Group I and Group II are inconsistent with each other with a 99.98\% (3.5$\sigma$) confidence.  We also found a difference in the total binary frequency between these three groups (47\% in Orion vs. 29\% in Group II vs. 45\% in Group III).  

    In order to determine if these differences in the binary separation distributions could be due to a systematic difference in the age of these sub-samples, we compared the spectral index distributions of the objects in Orion and the objects in Group II (Figure 12).  The median spectral index (as described in Section 5.1) for the targets in Orion is +0.72 and the median for Group II is +0.47, thus the objects in Orion appear to be slightly less evolved on average.  The Kolmogorov-Smirnov test showed that these spectral index distributions are inconsistent with being drawn from the same population with a 98.2\% (2.36~$\sigma$) confidence.  Within 3~$\sigma$ confidence limits these distributions are not statistically distinct, and thus it is unlikely that the difference between the binary separation distributions is due to age or evolutionary status.  A striking difference between the binary separation distributions for the sources in Orion versus the sources in Group II is the large excess of widely separated binary companions among the Orion sources.  In the comparison of the binary separation distribution versus spectral index with the whole sample, such a large difference in the binary frequency at wide separations occurred over a much larger change in spectral index (from $\alpha\approx1.4$ to 0.1, see Figures 4 and 5).  The fact that we see a similarly large difference in the binary frequency at wide separations over a much smaller difference in spectral index ($\Delta\alpha=0.25$ versus 1.3), as well as the large difference in binary frequency at separations closer than $\sim300$~AU, further suggest that the cause of the difference in these binary separation distributions is not due to age or evolutionary status. 

    There are several environmental factors that can affect the formation and survival of binary stars: first among them are star density, cloud temperature, and turbulence.  Although it is well known that the stellar density of a star forming cluster affects the long term survival of binaries in the cluster \citep{Pat2001, Kro1995b}, we do not believe that this is an important effect for our Class I YSOs.  None of our objects are in dense cluster forming environments, and the time scale for the tidal stripping of loosely bound companions is on the order of the cluster crossing time \citep{Kro2001} and is much greater than the Class I life time.   Although the stellar density around our Class I YSOs is not important, the difference between the binary separation distributions for Orion and Group II may be linked to the fact that the Orion cloud is forming a cluster with massive stars while the clouds hosting the sources in Group II are not.  Unfortunately, it is unclear which cloud property, if any, is responsible for the observed difference in the Class I binary separation distributions.  Since YSOs have experienced dynamical evolution before the Class I phase, the Class I binary properties may not be closely correlated with the pre-stellar cloud properties.
   
\section{Summary and Conclusions}
    We summarize the results of our research as follows:
    
      1)   We found a clear excess of binary stars among Class I YSOs over solar-type main sequence stars, especially at wide separations.  This result is consistent with the hypothesis that the vast majority of stars, and possibly all stars, form in binary and multiple groups.  However, we cannot estimate the total Class I binary frequency (integrated over all separations) since we have shown that a log-normal functional fit, used for other samples of binaries, is not applicable to Class I YSOs.  Thus, we cannot extrapolate the Class I binary separation distribution to a separation range where we do not have data.
      
      2) The large number of widely separated companion stars early in the Class I phase and the dearth of resolvable non-hierarchical multiple systems are consistent with the theoretical expectation that most binary stars form in non-hierarchical multiple systems that dynamically decay during the Class 0 phase.  
   
     3)  We observed that the binary frequency from 50~AU to 4500~AU is $\sim55\%$ for most of the Class I phase, then rapidly drops at the end of the Class I phase to $\sim15\%$.  The binary separation distribution also changes significantly within the Class I phase.
               
     4) It is clear that dynamical evolution is not over at the end of the Class 0 phase.  We propose a scenario that explains the decline in the binary frequency at separations wider than $\sim$1000~AU with decreasing spectral index using ejections within a dissipating envelope.  An increase in the binary frequency at separations near the expected formation radius for disk fragmentation early in the Class I phase, as well as the stability in the overall binary frequency despite the loss of binary companions at wide separations, suggest that new companion stars may be formed in the Class I phase.  
     
     5) The Class I binary separation distribution has a strong dependence on the star forming region.  Specifically, the binary separation distribution in Orion is different than the binary separation distribution in nearby, low mass star forming regions with a 3.5~$\sigma$ confidence.  The stellar density of the star forming environment is not relevant in our case, and it is unclear what cloud properties are the cause of this difference.  We anticipate that future theoretical work will explain the dependence of the binary statistics on the cloud properties.

\emph{Acknowledgments}  We thank the anonymous referee for useful comments.  This research has made use of the SIMBAD database, operated at CDS, Strasbourg, France.  This research has made use of NASA's Astrophysics Data System.  This publication makes use of data products from the Two Micron All Sky Survey, which is a joint project of the University of Massachusetts and the Infrared Processing and Analysis Center/California Institute of Technology, funded by the National Aeronautics and Space Administration and the National Science Foundation.  This research was supported by an appointment to the NASA Postdoctoral Program at the Ames Research Center, administered by the Oak Ridge Associated Universities through a contract with NASA. B.R. was supported by the NASA Astrobiology Institute under Cooperative Agreement No. NNA04CC08A and by the NSF through grants AST-0507784 and AST-0407005.


\begin{figure}
\plotone{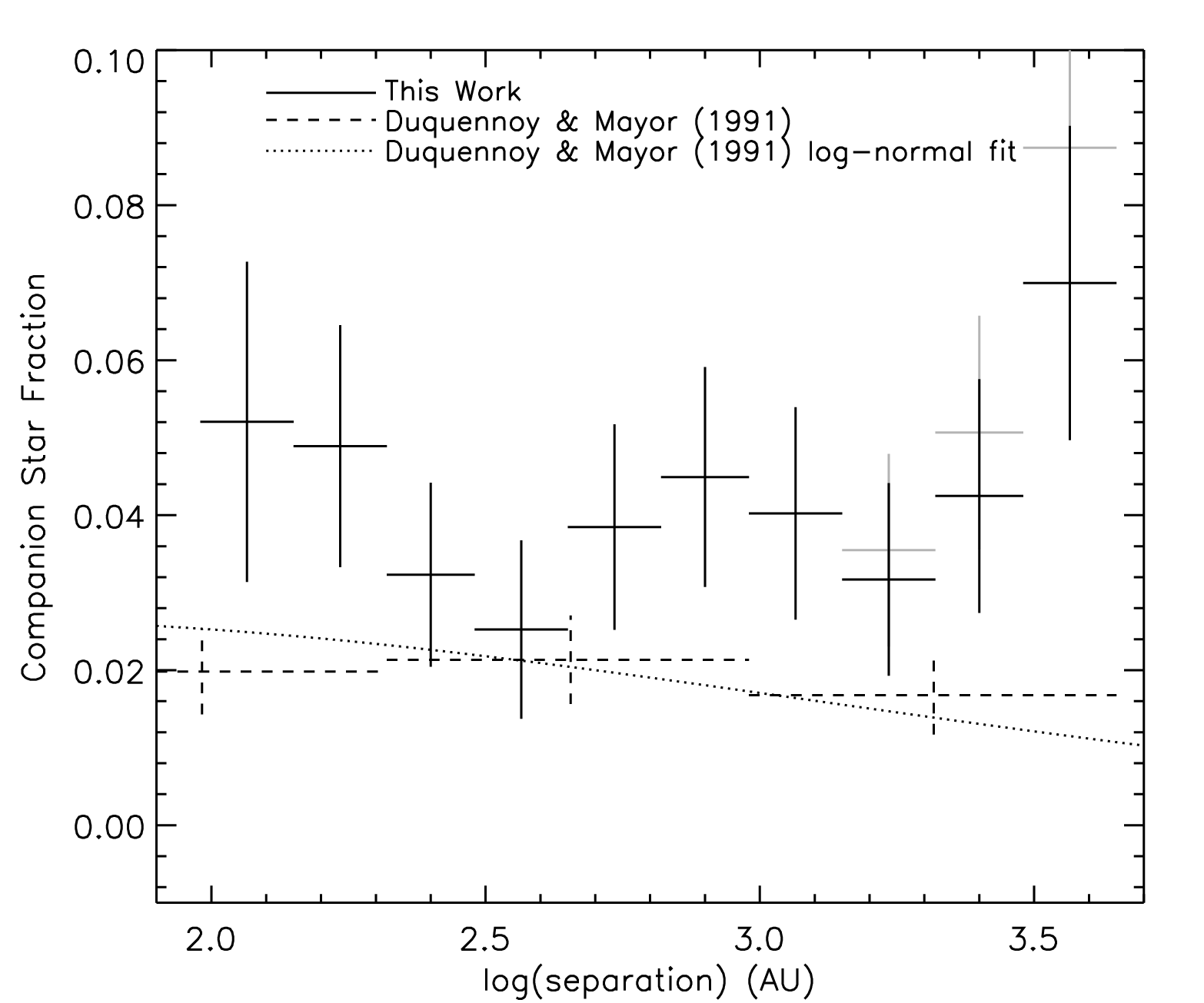}
\caption{The Class I binary separation distribution of our whole sample overlaid on the solar-type main sequence \citep{Duq1991} and T Tauri \citep{Pat2002} binary separation distributions.  The black crosses are our data.  The gray crosses are the values for the three outer most separation bins before we applied the correction for contamination.  The dashed crosses are the data from \citet{Duq1991}, and the dotted curve is their log-normal fit to their data.  Class I objects have a greater binary frequency than solar-type main sequence stars at separations wider than $\sim$100~AU.}
\end{figure}

\begin{figure}
\plotone{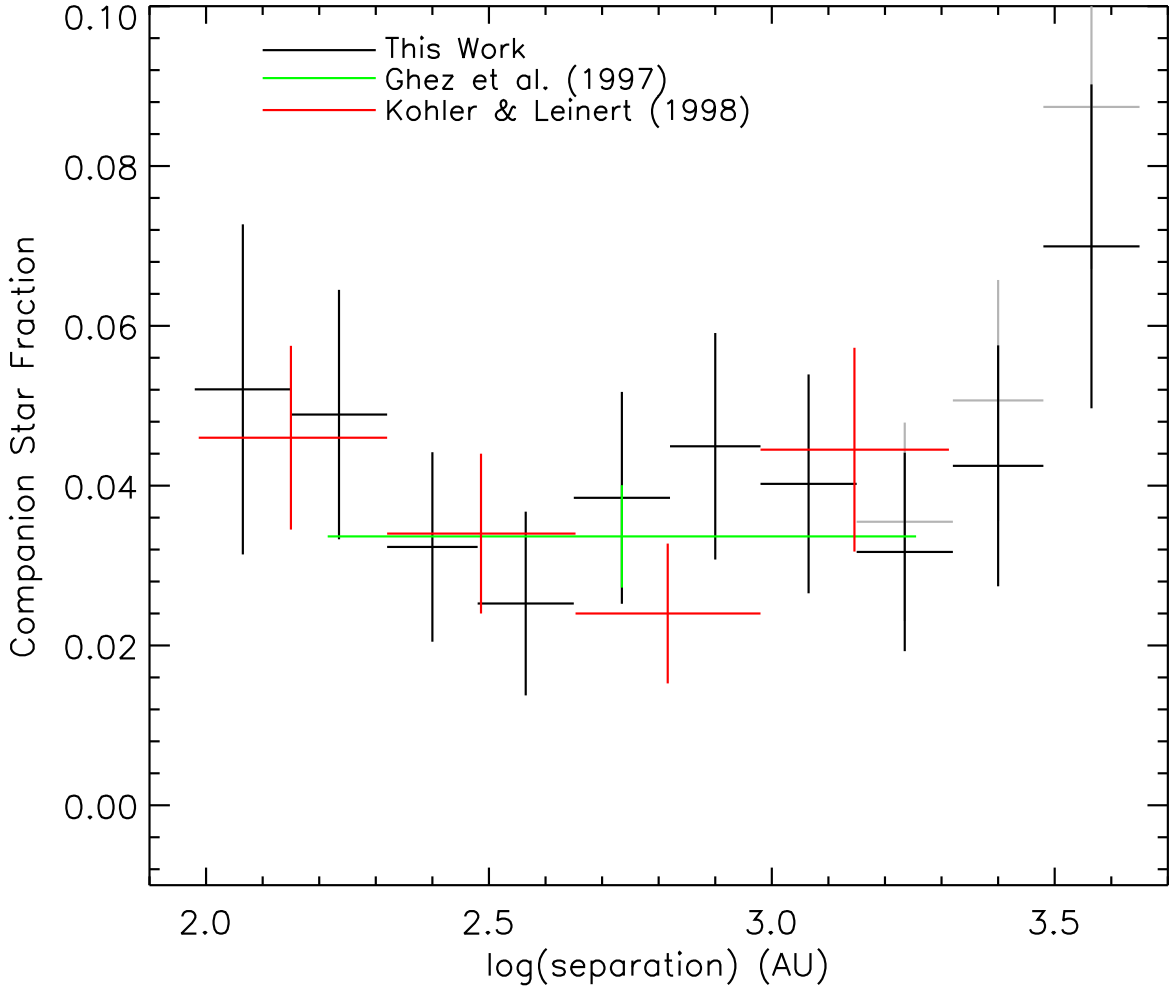}
\caption{The Class I binary separation distribution of our whole sample overlaid on T Tauri binary frequency measurements from \citet{Ghe1997} (in green) and \citet{Koh1998} (in red), scaled to match our bin size.  The Class I and T Tauri binary frequency distributions are consistent with each other.}
\end{figure}

\begin{figure}
\plotone{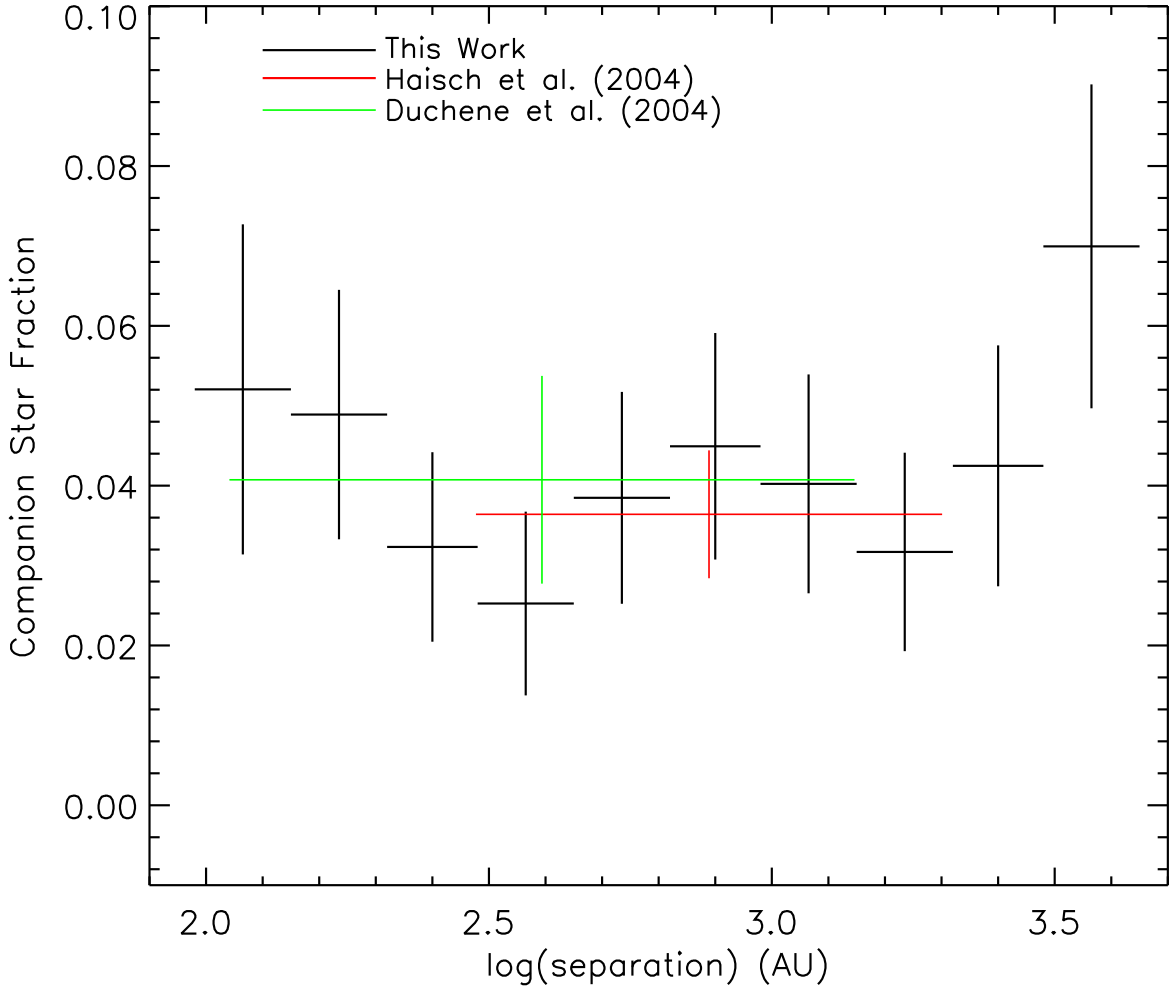}
\caption{The Class I binary separation distribution of our whole sample overlaid on the Class I binary frequency measurements from \citet{Duc2004} (in green) and \citet{Hai2004} (in red), scaled to match our bin size.  The results of these three studies agree very well.}
\end{figure}

\begin{figure}
\plottwo{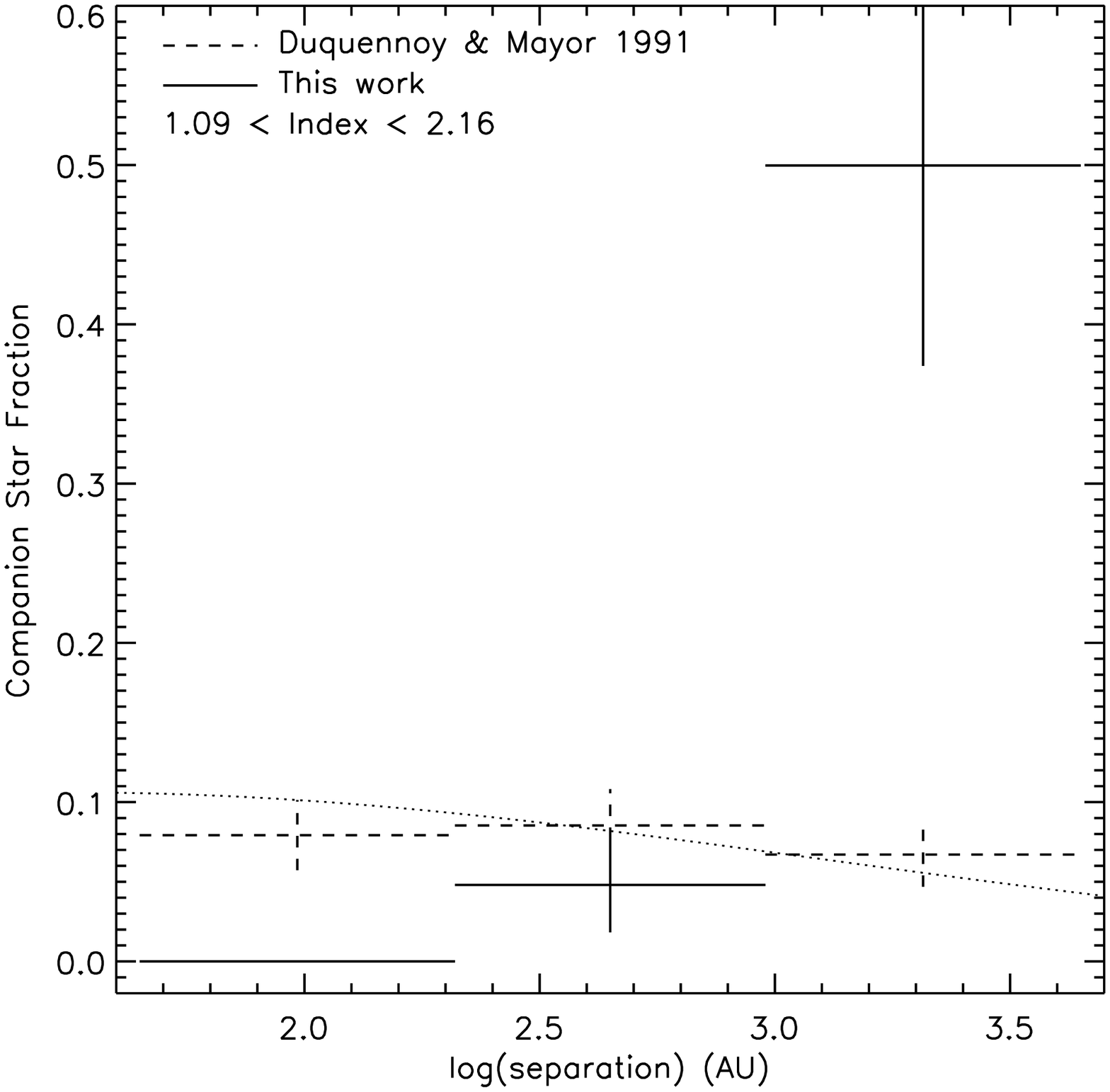}{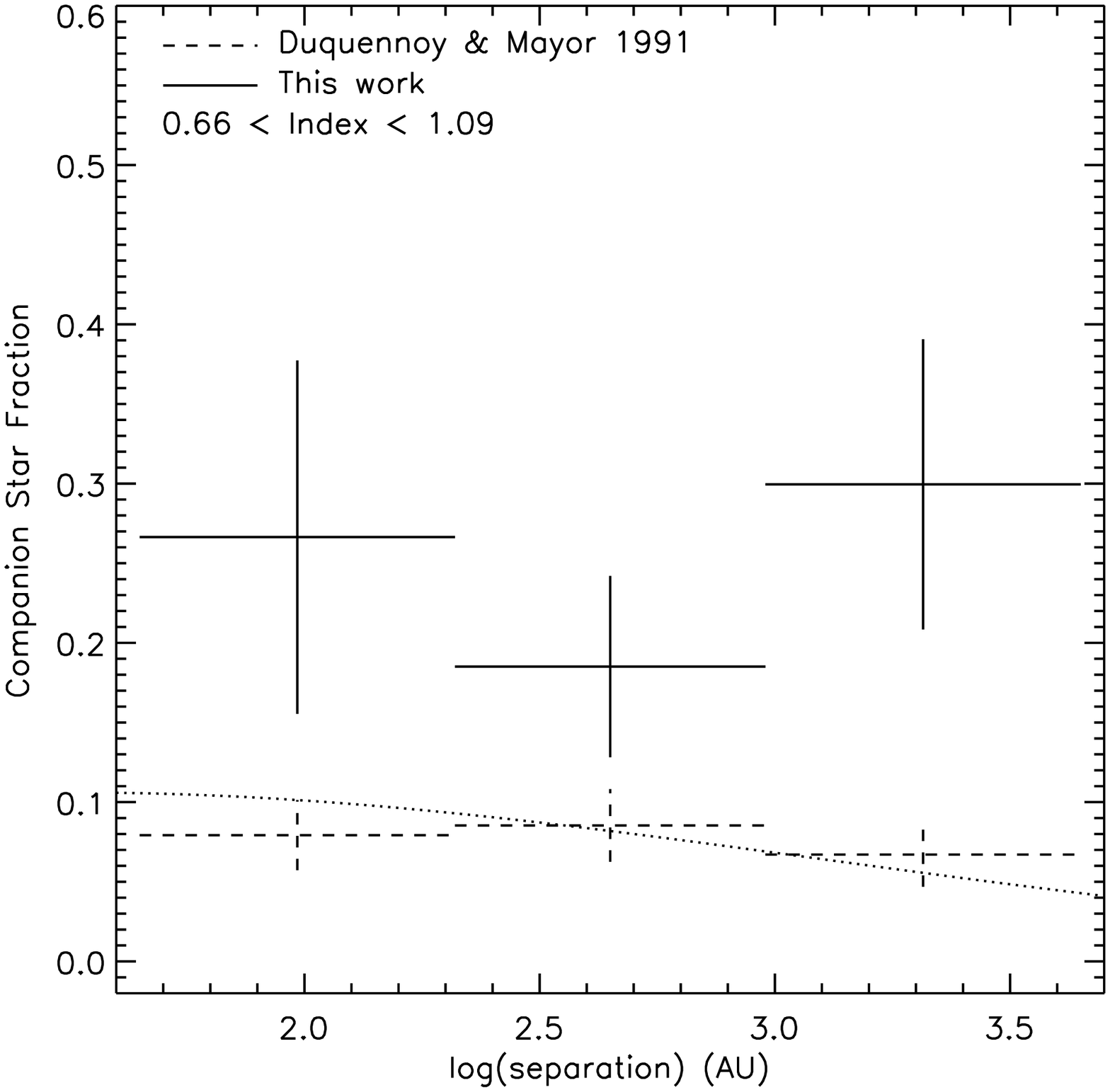}
\caption{The binary separation distributions for $1.09<\alpha<2.16$ and $0.66<\alpha<1.09$.  The figure on the left is the least evolved group in our sample, and remarkably nearly all of the resolvable binary companions were found at separations greater than $\sim$1000~AU.  For the second spectral index group (right), the binary frequency at wide separations has declined, while there are now several companions at separations less than $\sim$200~AU.}
\end{figure}

\begin{figure}
\plottwo{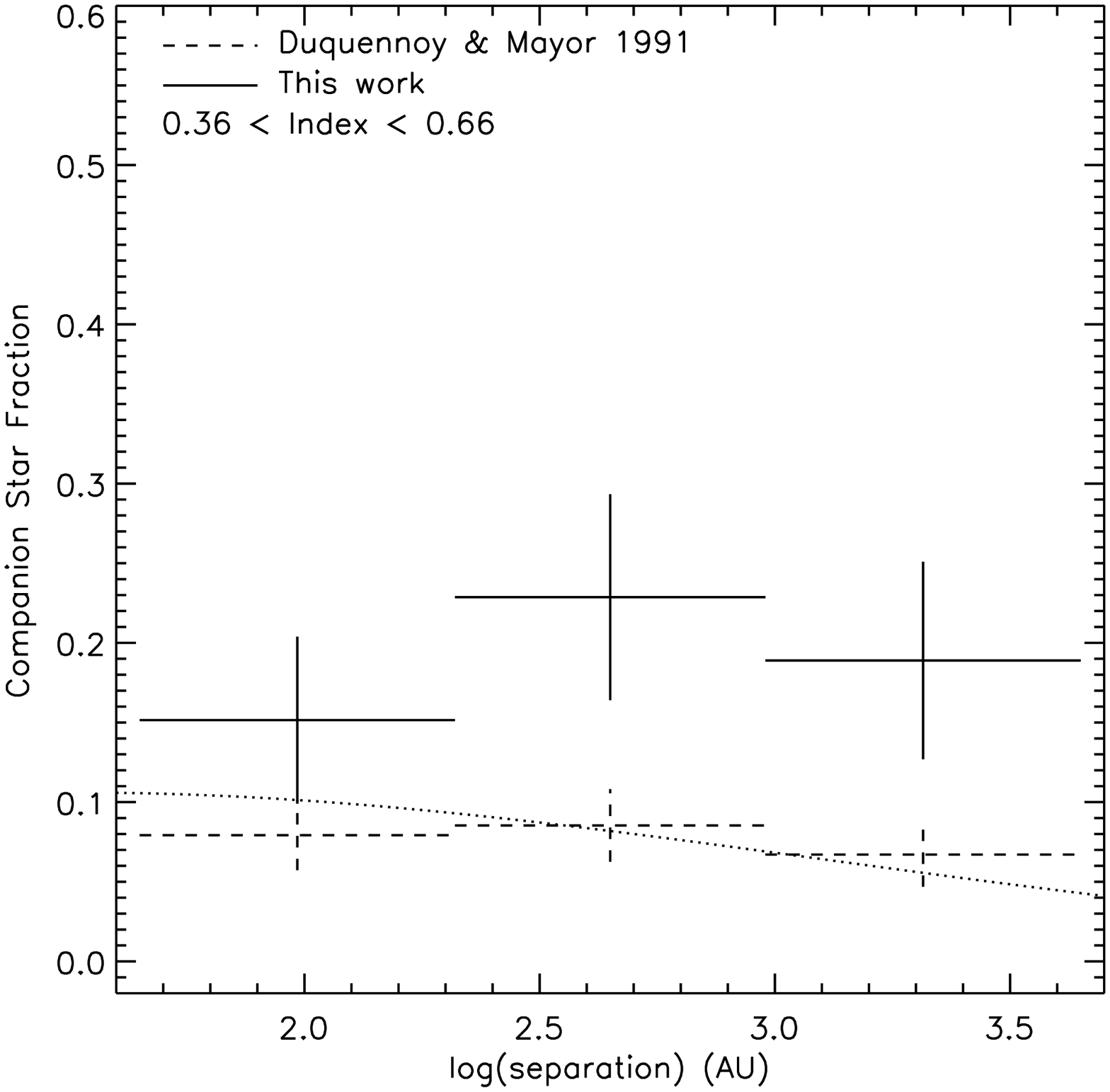}{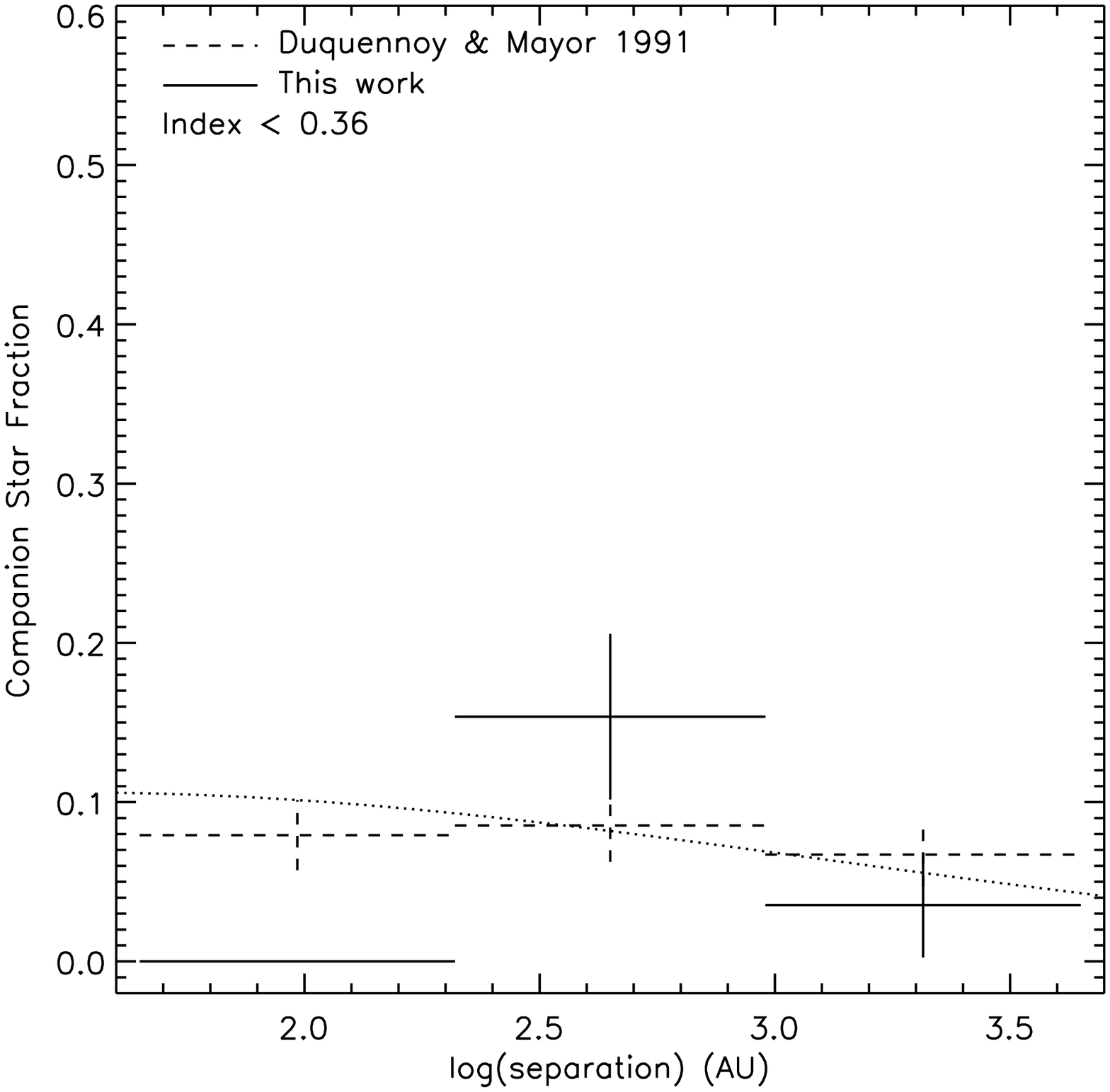}
\caption{The binary separation distributions for $0.36<\alpha<0.64$ and $\alpha<0.36$. In the third spectral index group (left), the binary frequency at intermediate separations is higher than before, while it is again lower at wide separations.  For the most evolved spectral index group (right), the binary frequency is lower at all separations.}
\end{figure}

\begin{figure}
\plotone{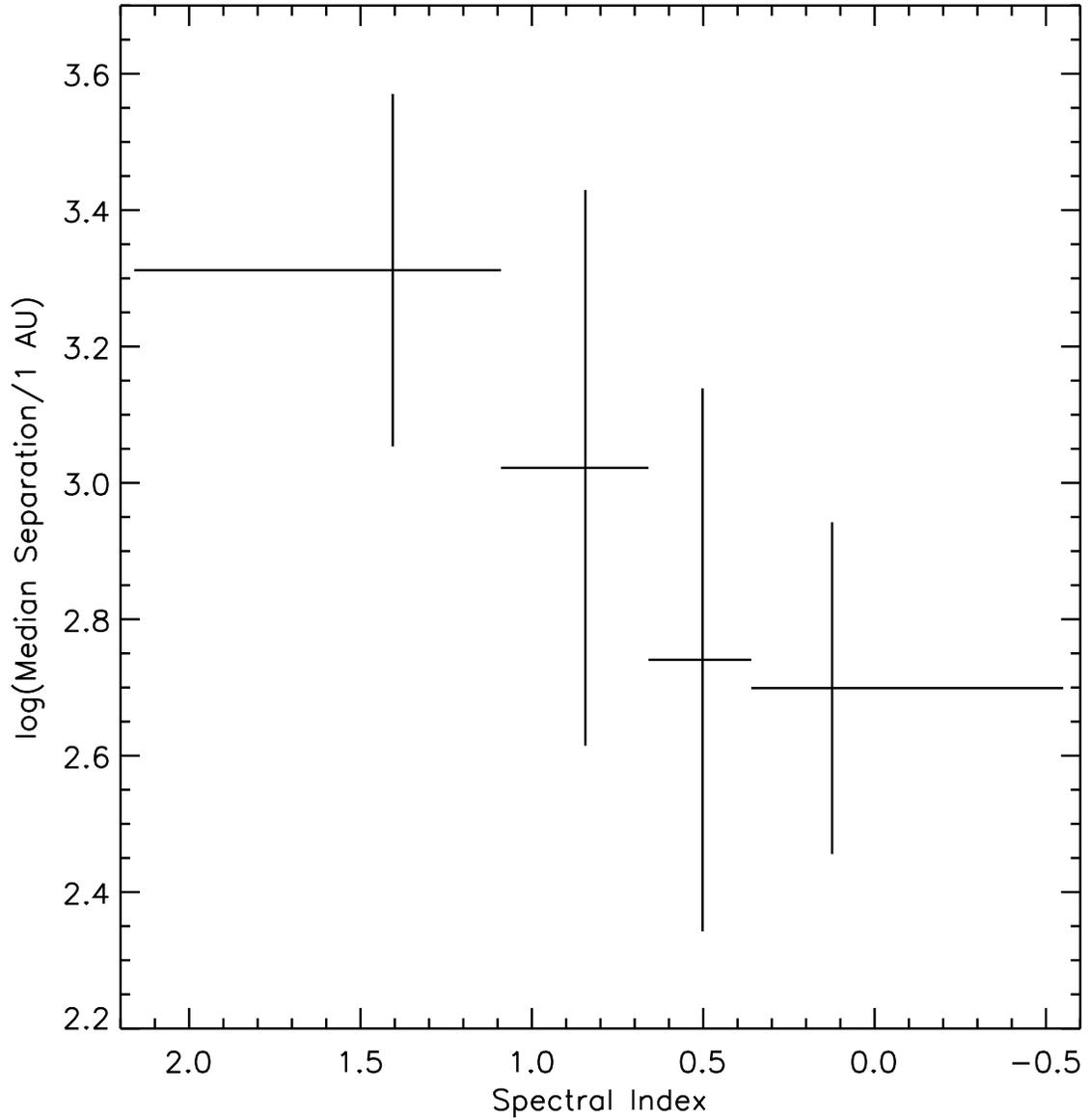}
\caption{The evolution of the median binary separation versus decreasing spectral index.  As before, the horizontal bar spans the range of spectral index included in that data point, and the error bar is at the location of the median spectral index of that data set.  In this case, the error bars are the median standard deviation (described in the text) of the log of the binary separations in each spectral index domain.  The median binary separation is observed to decrease with decreasing spectral index, although the binaries in each spectral index domain have a wide range of separations.}
\end{figure}

\begin{figure}
\plotone{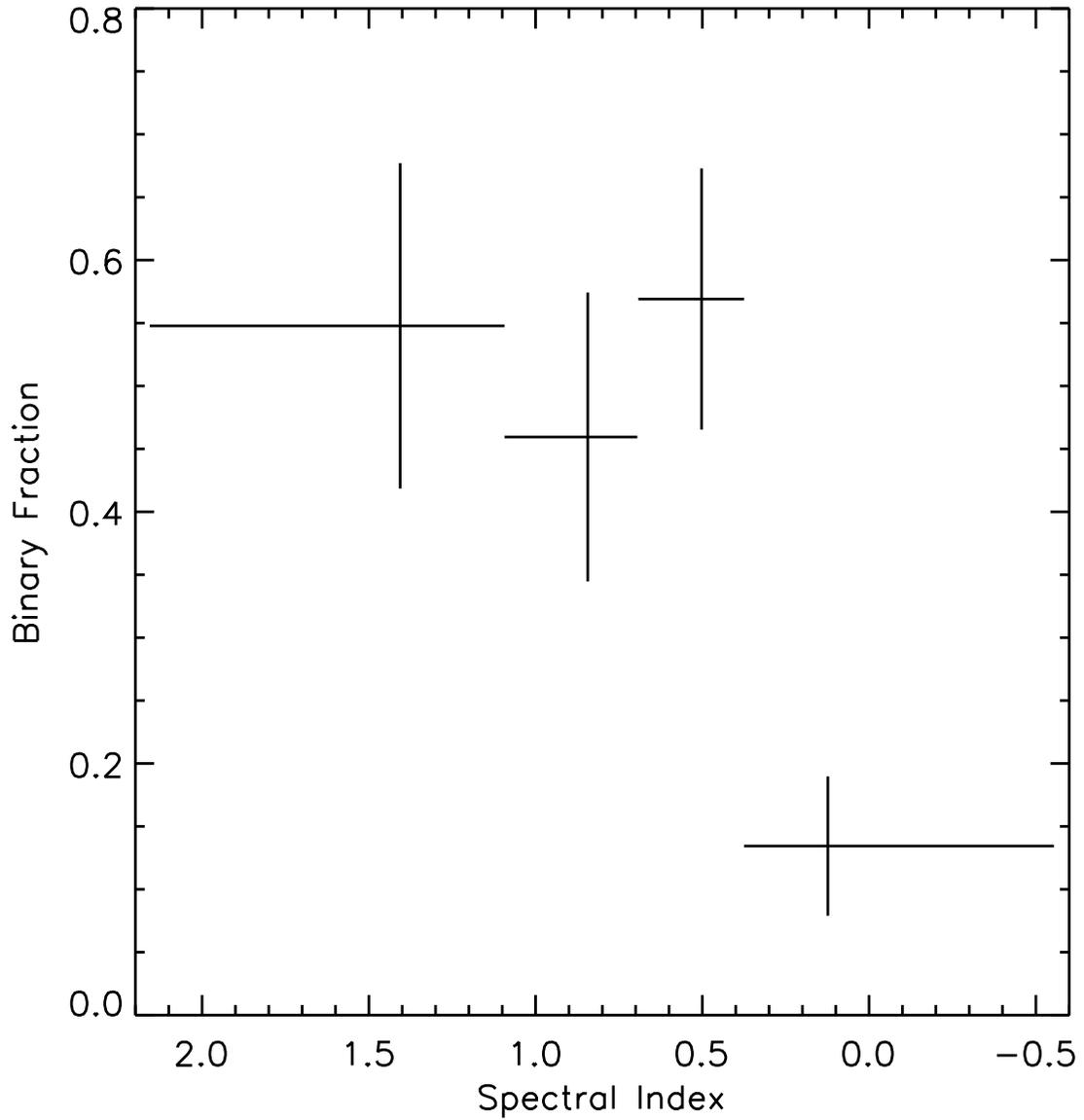}
\caption{The evolution of the binary frequency versus spectral index, including companion stars in the separation range from 50~AU to 4500~AU.  For each data point, the horizontal bar spans the range of spectral index included in that data point, and the error bar is at the location of the median spectral index of that data set.  The binary frequency is constant for most of the Class I phase, but is much lower in the most evolved spectral index group.}
\end{figure}

\begin{figure}
\plotone{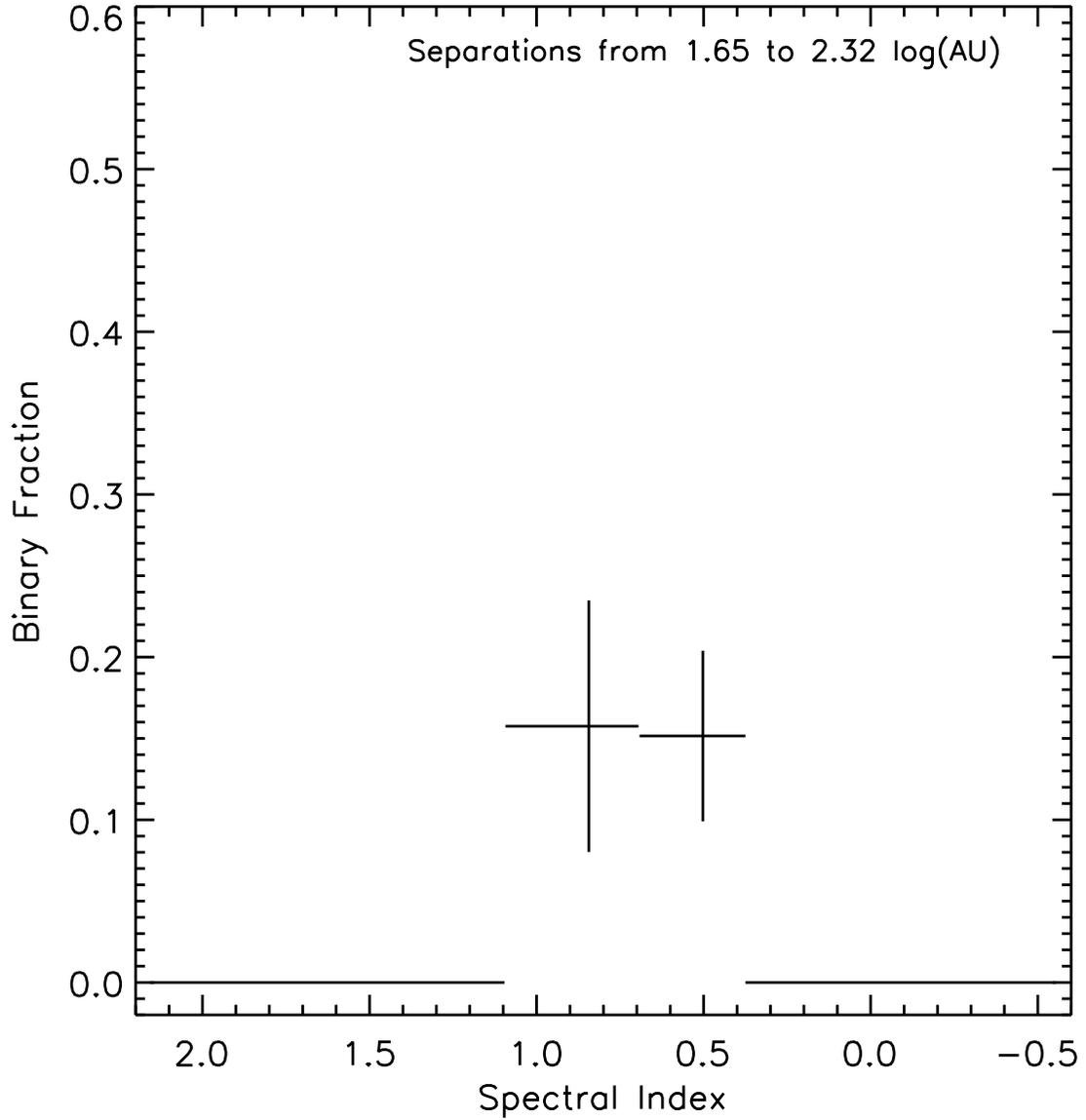}
\caption{The evolution of the binary frequency at separations from 45~AU to 207~AU (1.65 to 2.32 log(\emph{d}/1~AU) versus spectral index. In this separation range, the binary frequency is very low early in the Class I phase, rises, then declines again to a very low value.}
\end{figure}

\begin{figure}
\plotone{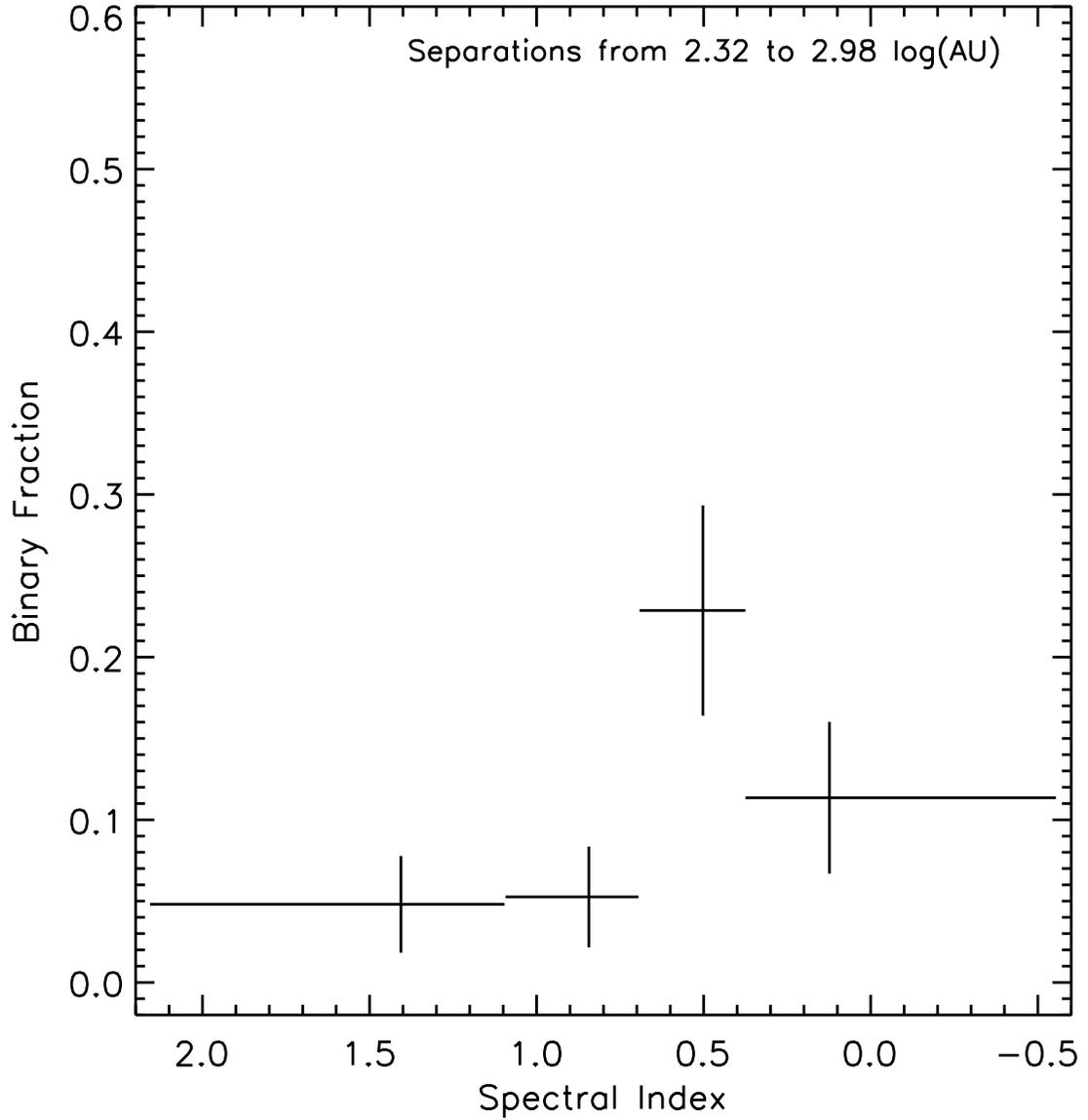}
\caption{The evolution of the binary frequency at separations from 207~AU to 963~AU (2.32 to 2.98 log(\emph{d}/1~AU) versus spectral index.  In this separation range, the binary frequency is low for the first two spectral index bins, increases sharply, then declines the end of the Class I phase.}
\end{figure}

\begin{figure}
\plotone{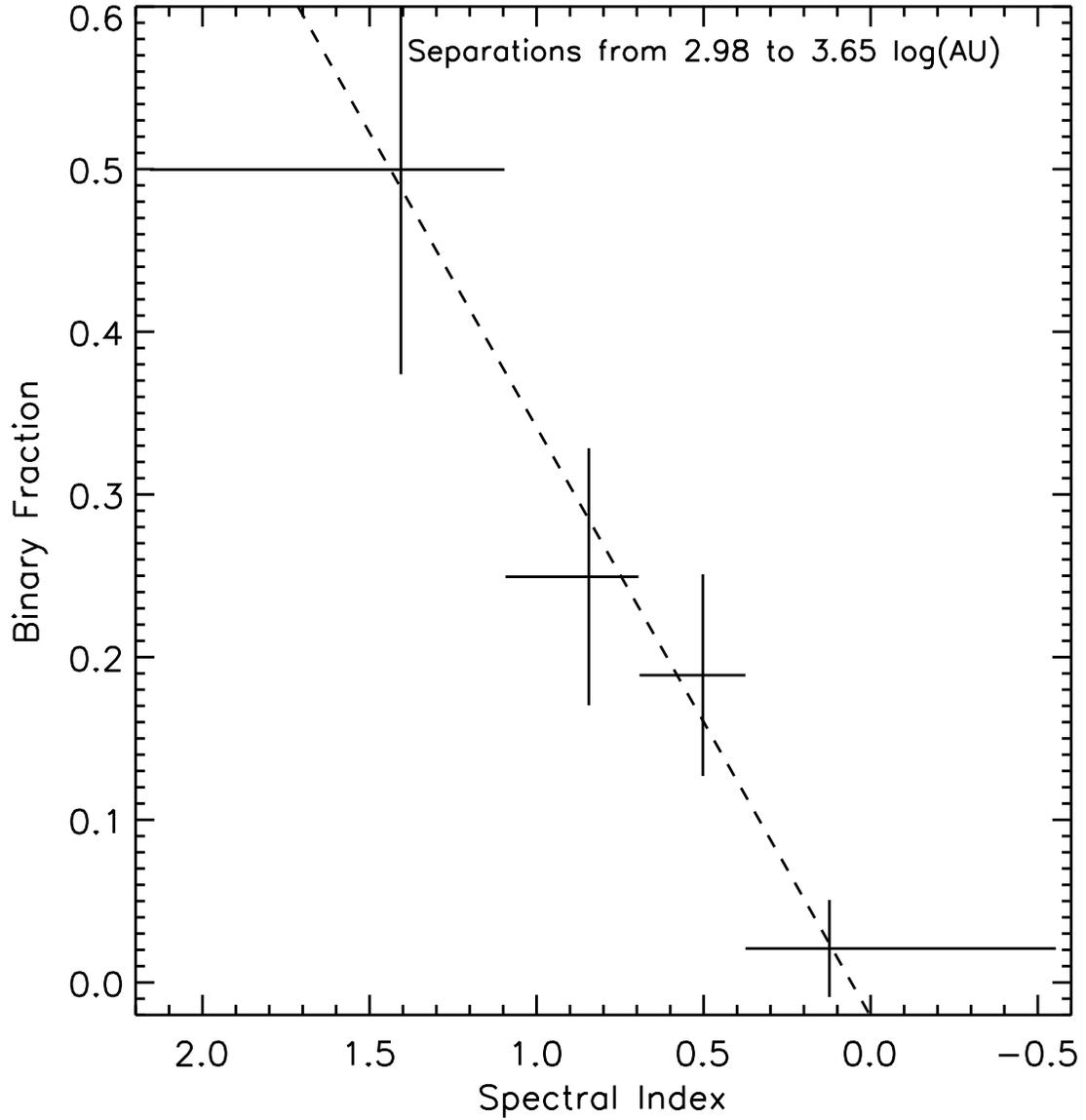}
\caption{The evolution of the binary frequency at separations from 963~AU to 4469~AU (2.98 to 3.65 log(\emph{d}/1~AU)) versus spectral index.  The dotted line is a linear fit to the data, showing that the decline of the binary frequency at wide separations with respect to spectral index is remarkably steady.}
\end{figure}

\begin{figure}
\plottwo{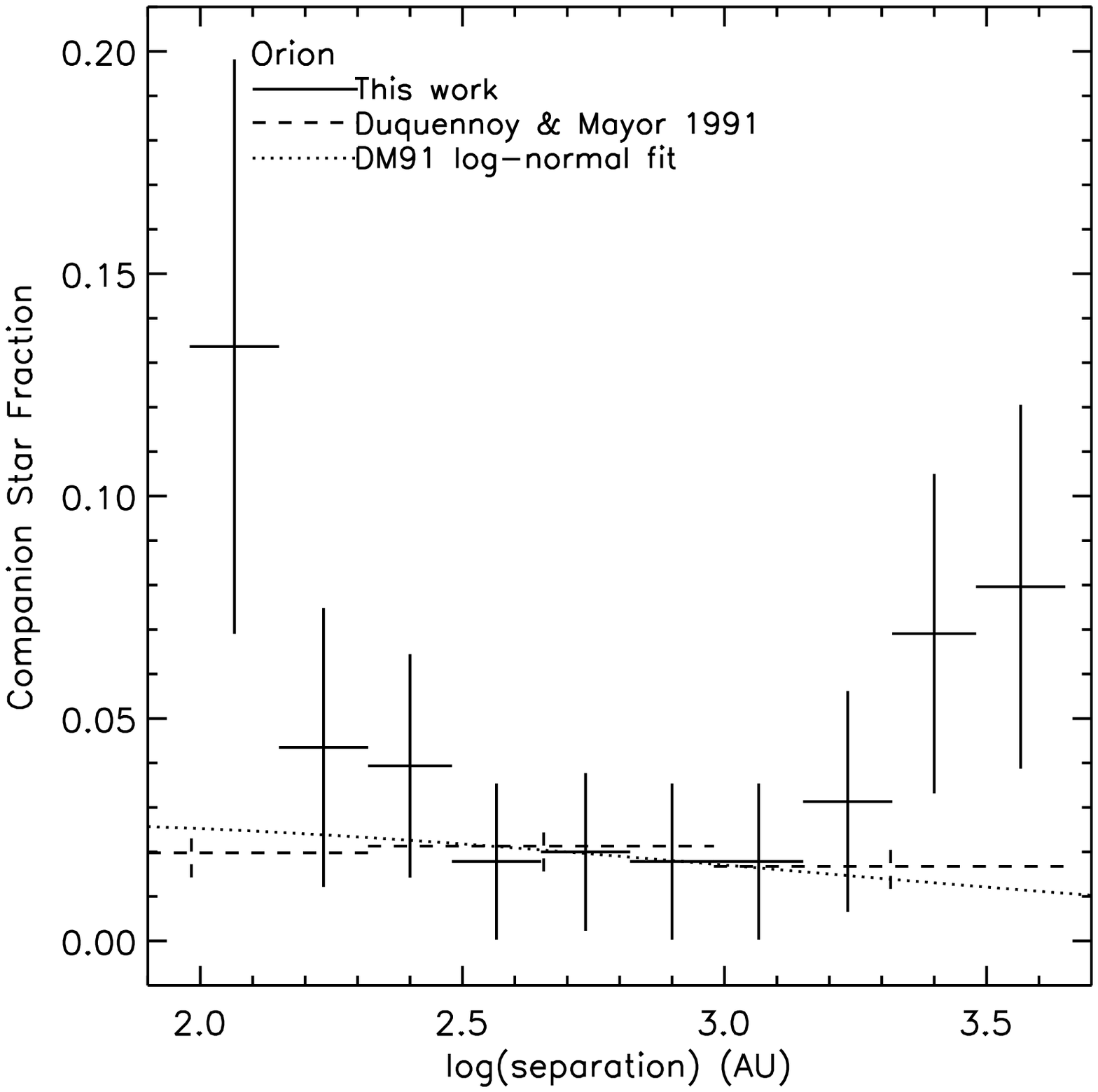}{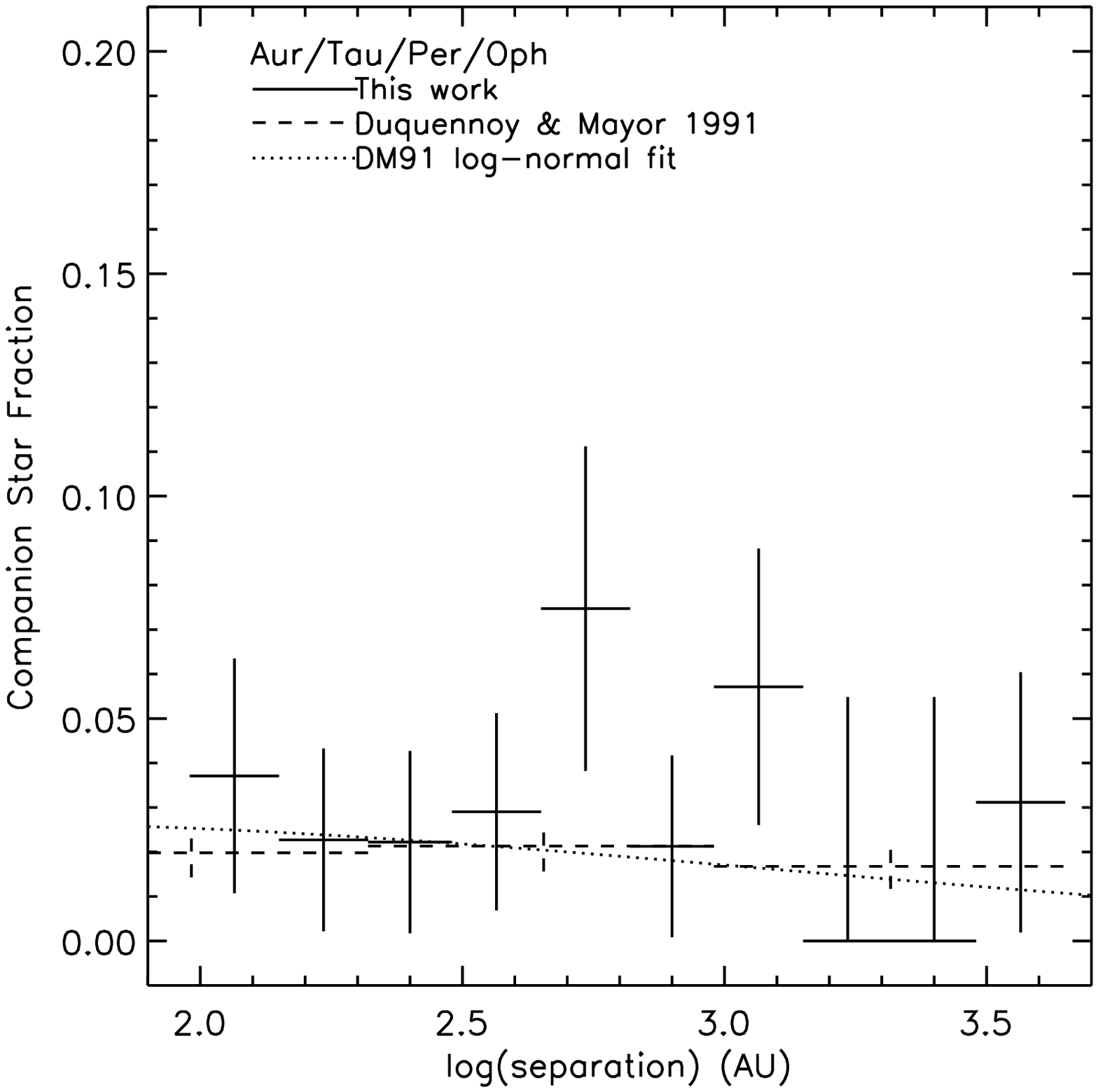}
\caption{The binary separation distributions in Group I (left) and in Group II (right).  The Class I objects in Group I have a binary excess at very close and wide separations; at intermediate separations the binary frequency is consistent with main sequence values.  The Class I sources in Group II have a binary frequency excess at intermediate separations; at close and wide separations the binary frequency is consistent with main sequence values.  These two groups are not likely to be drawn from the same parent population.}
\end{figure}

\begin{figure}
\plottwo{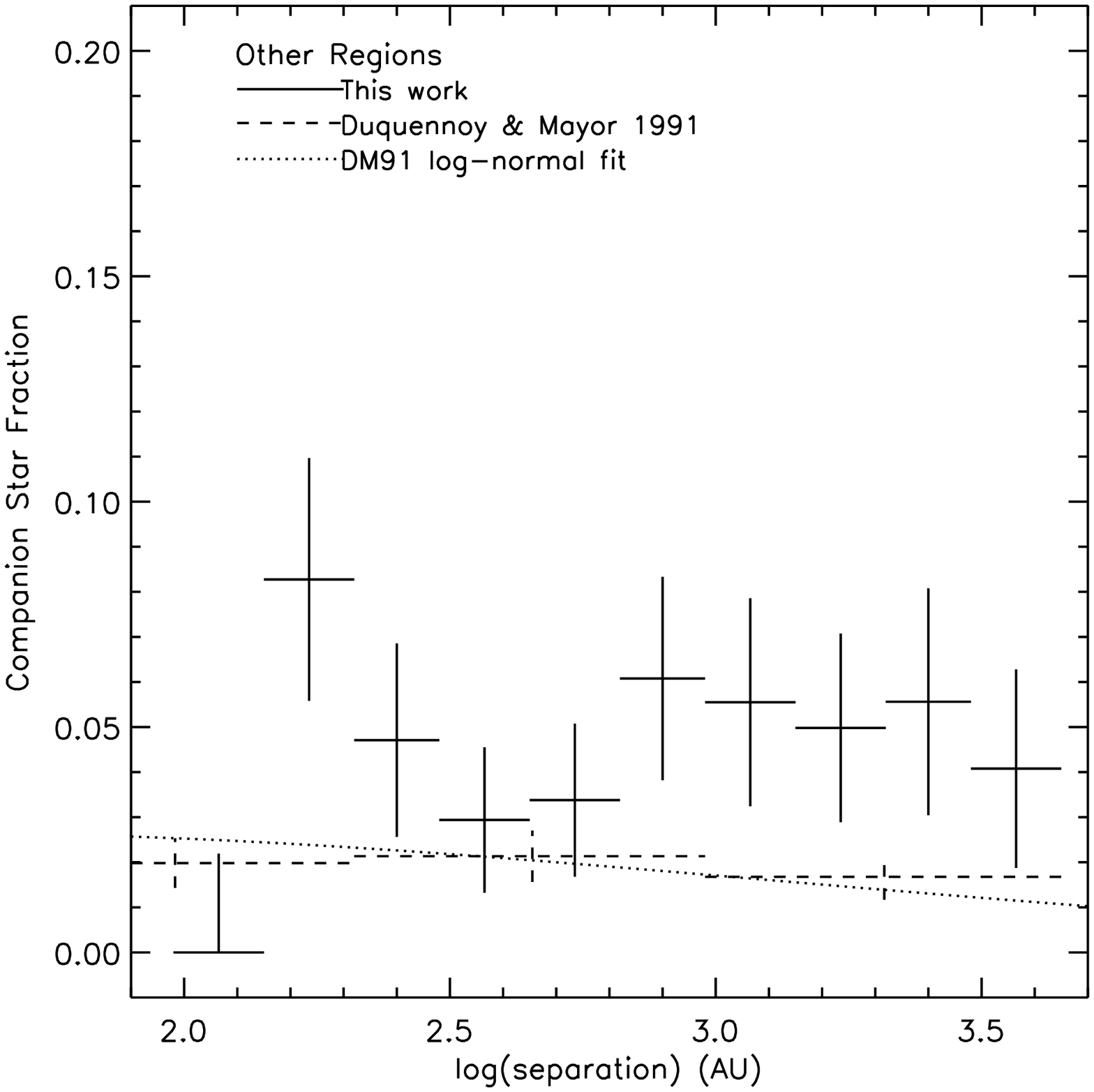}{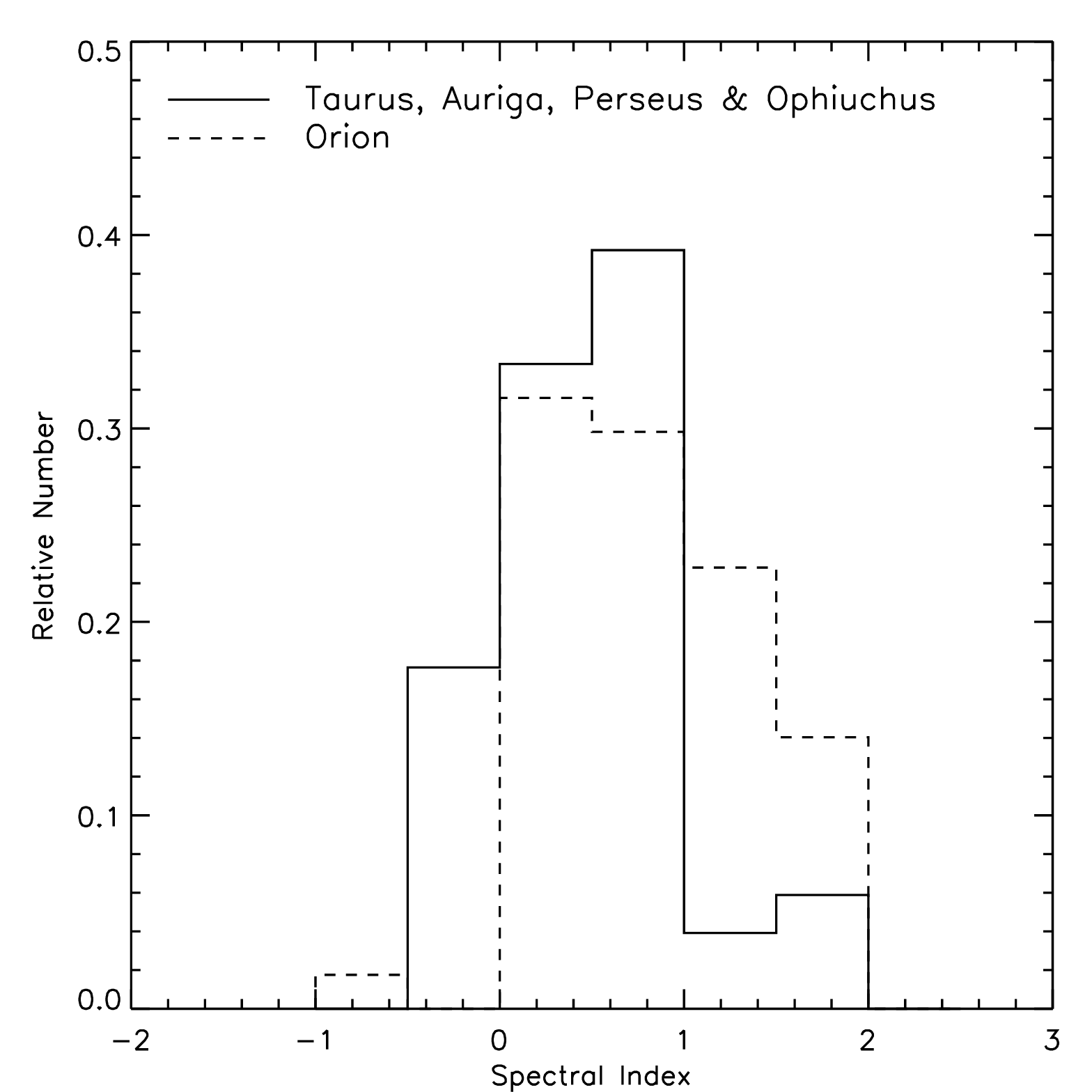}
\caption{The binary separation distribution for Class I objects in Group III (left) is very similar to the binary separation distribution for our whole sample.  The spectral index distributions for our objects in Orion and in Taurus/Auriga/Perseus/Ophiuchus (right) are also quite similar.  The Kolmogorov-Smirnov test shows that these distributions have different parent populations with less than a 3$\sigma$ confidence.  This figure has been corrected for contaminaton.}
\end{figure}

\end{document}